\pdfoutput=1
\documentclass[sigconf,nonacm]{acmart}
\setcopyright{none}

\renewcommand\footnotetextcopyrightpermission[1]{}

\PassOptionsToPackage{protrusion, expansion}{microtype} %

\usepackage{amsmath, amsthm}

\usepackage{graphicx, xcolor} %
\usepackage{hyperref, url}
\usepackage{array}
\usepackage{pifont}
\usepackage{rotating}

\newcommand{\cmark}{\ding{51}}%
\newcommand{\xmark}{\ding{55}}%
\definecolor{crimson}{rgb}{0.86, 0.08, 0.24}
\definecolor{greenarrow}{RGB}{57, 151, 92}
\usepackage{booktabs, multirow}
\usepackage{caption}
\usepackage{subcaption}
\usepackage{enumitem}
\usepackage{setspace}
\usepackage{xcolor}
\usepackage[page]{appendix}

\usepackage[ruled, linesnumbered, norelsize]{algorithm2e}

\usepackage{lmodern}
\usepackage{helvet}

\newcommand{\thiswork}{STZ}

\copyrightyear{2025}
\acmYear{2025}
\setcopyright{cc}
\setcctype{by}
\acmConference[SC '25]{The International Conference for High Performance Computing, Networking, Storage and Analysis}{November 16--21, 2025}{St Louis, MO, USA}
\acmBooktitle{The International Conference for High Performance Computing, Networking, Storage and Analysis (SC '25), November 16--21, 2025, St Louis, MO, USA}
\acmDOI{10.1145/3712285.3759795}
\acmISBN{979-8-4007-1466-5/2025/11}

\begin{document}

\title{\thiswork: A High Quality and High Speed Streaming Lossy Compression
Framework for Scientific Data}

\settopmatter{authorsperrow=4}

\author{Daoce Wang}
\affiliation{
  \institution{Univ of Nebraska Omaha}
  \city{Omaha}\state{NE}
  \country{USA}
}
\email{daocewang@unomaha.edu}

\author{Pascal Grosset}
\affiliation{
  \institution{Los Alamos National Lab}
  \city{Los Alamos}\state{NM}
  \country{USA}
}
\email{pascalgrosset@lanl.gov}

\author{Jesus Pulido}
\affiliation{%
  \institution{Los Alamos National Lab}
  \city{Los Alamos}\state{NM}
  \country{USA}
}
\email{pulido@lanl.gov}

\author{Jiannan Tian}
\affiliation{
  \institution{Oakland University}
  \city{Rochester}\state{MI}
  \country{USA}
}
\email{jtian@oakland.edu}

\author{Tushar M. Athawale}
\affiliation{%
  \institution{Oak Ridge National Lab}
  \city{Oak Ridge}\state{TN}
  \country{USA}
}
\email{athawaletm@ornl.gov}

\author{Jinda Jia}
\affiliation{
  \institution{Indiana University}
  \city{Bloomington}\state{IN}
  \country{USA}
}
\email{jindjia@iu.edu}

\author{Baixi Sun}
\affiliation{
  \institution{Indiana University}
  \city{Bloomington}\state{IN}
  \country{USA}
}
\email{sunbaix@iu.edu}

\author{Boyuan Zhang}
\affiliation{
  \institution{Indiana University}
  \city{Bloomington}\state{IN}
  \country{USA}
}
\email{bozhan@iu.edu}

\author{Sian Jin}
\affiliation{%
  \institution{Temple University}
  \city{Philadelphia}\state{PA}
  \country{USA}
}
\email{sian.jin@temple.edu}

\author{Kai Zhao}
\affiliation{%
  \institution{Florida State University}
  \city{Tallahassee}\state{FL}
  \country{USA}
}
\email{kai.zhao@fsu.edu}

\author{James Ahrens}
\affiliation{
  \institution{Los Alamos National Lab}
  \city{Los Alamos}\state{NM}
  \country{USA}
}
\email{ahrens@lanl.gov}

\author{Fengguang Song}
\authornote{Corresponding author}
\affiliation{
  \institution{Indiana University}
  \city{Bloomington}\state{IN}
  \country{USA}
}
\email{fgsong@iu.edu}

\renewcommand{\shortauthors}{Wang et al.}

\begin{abstract}
  Error-bounded lossy compression is one of the most efficient solutions to reduce the volume of scientific data. For lossy compression, progressive decompression and random-access decompression are critical features that enable on-demand data access and flexible analysis workflows. However, these features can severely degrade compression quality and speed.
  To address these limitations, we propose a novel streaming compression framework that supports \textit{both progressive decompression and random-access decompression} while maintaining high compression quality and speed. Our contributions are three-fold: (1) we design the first compression framework that simultaneously enables both progressive decompression and random-access decompression; (2) we introduce a hierarchical partitioning strategy to enable both streaming features, along with a hierarchical prediction mechanism that mitigates the impact of partitioning and achieves high compression quality---even comparable to state-of-the-art (SOTA) non-streaming compressor SZ3; and (3) our framework delivers high compression and decompression speed, up to 6.7$\times$ faster than SZ3.

\end{abstract}

\keywords{Lossy compression, progressive decompression, random-access.}

\maketitle

\setlength{\textfloatsep}{6pt}
\setlength\abovecaptionskip{3pt}

\newpage
\begingroup
\renewcommand{\contentsname}{\Large\textbf{Contents}}
\renewcommand{\addcontentsline}[3]{}%
\tableofcontents
\endgroup
\newpage

\section{Introduction}

Supercomputers have quickly approached exascale capabilities with increasingly higher computing power, causing the volume of scientific data storage and transmission to grow exponentially. Traditionally, a simulation with a resolution of \(4096^3\) can generate approximately 5~TB of data per snapshot. Thus, running a single simulation with 200 snapshots would require a total disk storage capacity of 1~PB. Simulations in exascale scenarios are even larger---often involving thousands of points per axis---highlighting the necessity for effective data size reduction. While computing power has advanced rapidly, storage bandwidth has not kept pace, widening the gap between processing capability and data movement.

To fill this gap, data reduction techniques have become critical for conserving storage bandwidth, reducing power consumption, and optimizing HPC system performance. Traditional lossless compression techniques, such as GZIP~\cite{gzip} and Zstandard~\cite{zstd}, typically achieve compression ratios of only up to 2 for scientific data. In contrast, new generations of lossy compressors---known as error-bounded lossy compressors---have been developed, including SZ~\cite{di2016fast, sz17, sz18, zhao2021optimizing}, ZFP~\cite{zfp}, MGARD~\cite{gong2023mgard, ainsworth2018multilevel}, SPERR~\cite{sperr}, and TTHRESH~\cite{ballester2019tthresh}. These compressors provide significantly higher compression ratios while minimizing quality loss in the reconstructed data and preserving its fidelity for post-analysis.

\begin{figure}[h]
  \centering
  \begin{subfigure}[t]{\linewidth}
    \centering
    \includegraphics[width=\linewidth]{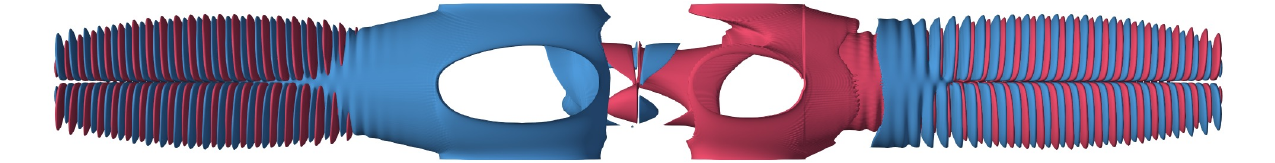}
    \label{fig:vis-wpx-fine}
  \end{subfigure}
  \vspace{2mm}
  \begin{subfigure}[t]{\linewidth}
    \centering
    \includegraphics[width=\linewidth]{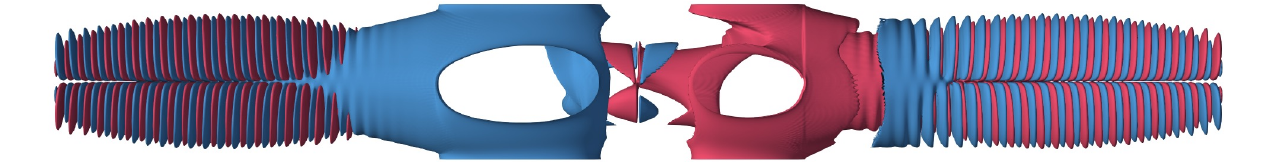}
    \label{fig:vis-wpx-coarse}
  \end{subfigure}~%
  \caption[t]{Iso-surface of original (top, $256\times256\times2048$, $1GB$) and down-sampled (bottom, $128\times128\times1024$, $128MB$) WarpX dataset. Visual differences between them are virtually imperceptible (SSIM=0.96), highlighting the need for progressive decompression.}
  \label{fig:vis-prog-why}
\end{figure}

Although lossy compression can significantly reduce the storage and transmission overhead, traditional compression workflows require the entire compressed data to be fully decompressed for post-analysis or visualization, which restores the dataset to its original uncompressed size.
However, many scientific use cases, such as inspecting regions of interest (e.g., 2D slices in a 3D dataset) or generating coarse-resolution previews~\cite{Progressive1,Progressive2} (as shown in Figure~\ref{fig:vis-prog-why}), do not require full data reconstruction.
In these scenarios, decompressing and storing the entire dataset introduces significant computational, memory, I/O, and storage overhead. \textit{This problem is especially severe in two cases:
  (1) For analyzing terabyte-scale datasets on local devices, limited RAM and disk space can severely restrict the user's ability to decompress and process the data.
(2) For larger datasets, such as petabyte-scale datasets from extreme-scale simulations, the decompressed data can be too large to fit even in the memory of one or a few HPC nodes, making full decompression impractical on HPC systems.}
These challenges highlight the urgent need for on-demand decompression frameworks that support targeted access to compressed scientific data without reconstructing the entire dataset.

Streaming compression \textcolor{black} {is a method} that supports both \textit{progressive} \textit{decompression} (i.e., decompressing data to a lower resolution) and \textit{random-access} \textit{decompression} (i.e., selectively decompressing parts of the data), offering an effective solution to these challenges.
However, many scientific compressors, such as SZ3, do not support streaming compression, as it can significantly degrade compression quality and speed.
Moreover, to the best of our knowledge, no existing scientific compressor supports both progressive decompression and random-access decompression simultaneously. For example, ZFP supports random-access decompression by partitioning data into \(4^3\) blocks and processes each block independently during compression. However, this block-wise approach sacrifices compression quality/ratio due to the loss of spatial information between blocks (as summarized in Table~\ref{tab:first}). On the other hand, SPERR supports progressive decompression through its wavelet transform, which enables lower-resolution extraction while preserving global data structure. However, the wavelet transform used by SPERR is computationally expensive, resulting in slow performance and making it unsuitable for in-situ \textcolor{black}{or online} compression scenarios.

\begin{table}[h]

  \vspace{-1mm}
  \centering\sffamily
  \caption{Features of different compressors.}
  \label{tab:first}
  \resizebox{0.97\linewidth}{!}{
    \begin{tabular}{@{}rccccc@{}}
      \toprule
      & SZ3 & SPERR & MGARD & ZFP & \textbf{Ours} \\
      \midrule
      Progressive   & \textcolor{red}{\xmark} &\textcolor{greenarrow}{\cmark} & \textcolor{greenarrow}{\cmark} & \textcolor{red}{\xmark} & \textcolor{greenarrow}{\cmark} \\
      Random Access & \textcolor{red}{\xmark} & \textcolor{red}{\xmark} & \textcolor{red}{\xmark} & \textcolor{greenarrow}{\cmark} & \textcolor{greenarrow}{\cmark}\\
      Speed         & \textcolor{gray}{\textbf{mid}} & \textcolor{red}{\textbf{very low}} &\textcolor{red}{\textbf{low}} & \textcolor{greenarrow}{\textbf{very high}} & \textcolor{greenarrow}{\textbf{high}} \\
      Comp. Quality   & \textcolor{greenarrow}{\textbf{high}} & \textcolor{greenarrow}{\textbf{very high}} & \textcolor{gray}{\textbf{mid}} & \textcolor{red}{\textbf{low}} & \textcolor{greenarrow}{\textbf{high}} \\
      \bottomrule
    \end{tabular}
  }
\end{table}

To address these limitations, we propose a novel streaming compression framework that fills the gap between streaming capabilities and high compression quality and speed. Our framework supports both progressive decompression and random-access decompression, enabling flexible HPC workflows for handling massive scientific data. We evaluate our solution against four state-of-the-art (SOTA) compressors and demonstrate that it enables streaming capabilities without compromising compression quality and speed. The features of our work and the baseline compressors are summarized in Table~\ref{tab:first}, and detailed evaluation results are presented in \S\ref{sec:eva}.

Our contributions are summarized as follows:
\begin{enumerate}[leftmargin=*, label=\arabic*.]
  \item We propose a first-of-its-kind streaming compression framework that simultaneously supports progressive decompression and random-access decompression.
  \item To enable streaming compression, we first introduce a hierarchical partitioning strategy. To mitigate compression quality loss caused by partitioning, we propose a hierarchical prediction method to preserve spatial information across different hierarchical levels, enabling our framework to achieve even comparable compression quality to the non-streaming compressor SZ3.
  \item Our framework achieves high compression and decompression speeds, with high parallel efficiency, and is up to 6.7$\times$ faster than SOTA compressors with the same compression quality and ratio.
  \item  We evaluate our framework against four SOTA scientific lossy compressors: SZ3, SPERR, MGARD-X, and ZFP, using four real-world scientific datasets. The results show that our solution achieves the best overall trade-off between compression quality and speed. Notably, none of the SOTA compressors support both progressive decompression and random-access decompression.
\end{enumerate}

\section{Related Work}
\subsection {Lossy Compression for Scientific Data}
Most scientific simulations use floating-point data. There are two main categories of floating-point data compression: lossless and lossy compression.
Lossless compression suffers from low compression ratios on scientific data (only up to 2$\times$~\cite{chen2024fcbench}), because even when floating-point values are numerically similar, their binary representations can differ significantly, making them difficult to be effectively compressed. In contrast, lossy compression offers much higher compression ratios by trading off some accuracy and better exploiting the redundancy in floating-point data.
Unlike traditional image lossy compressors (e.g., JPEG), which focus on image compression, a new generation of error-bounded lossy compressors (specifically for scientific data) has been developed in recent years. These include SZ, ZFP, MGARD, SPERR, and TTHRESH, and their GPU versions~\cite{tian2020cusz,tian2021optimizing, cuZFP, huang2024cuszp2, liu2024cusz}. These compressors have been widely used in the scientific community~\cite{jin2024concealing,Huebl2017,lu2018understanding,luo2019identifying,cappello2019use,amric, baker2019evaluating, di2024survey, wang2022tac, wang2024tac+, wang2023amrvis, Jia2024SDP4Bit, sun2025compso, jia2024neurlz, wang2025designing, wang2024mrz, zheng2024vtk} as they offer higher compression quality for scientific datasets and offer strict error control based on user-specified bounds.

In general, there are three steps in error-bounded scientific lossy compressors:
(1) First, the data is de-correlated to leverage spatial redundancy, such as by using prediction (SZ3, MGARD) or transformation (ZFP, SPERR);
(2) Then, the loss is introduced to improve compressibility, for example via quantization (SZ3) or bit truncation (ZFP);
(3) Finally, the data goes through lossless encoding, such as Huffman encoding (SZ3, MGARD) or run-length encoding (TTHRESH).
Our work combines prediction, quantization, and Huffman encoding to achieve high speed and compression ratio.

\subsection {Progressive Decompression}
There are two main categories of progressive decompression: (1) precision-progressive and (2) resolution-progressive.

\textit{Precision-progressive decompression}~\cite{magri2023general, liang2021error, wu2024error, yang2025ipcomp} reconstructs data at multiple fidelity levels by incrementally increasing precision (i.e., reducing compression error). IPComp~\cite{yang2025ipcomp} is currently the best precision-progressive compressor in terms of overall compression quality and speed; however, its quality and speed are still lower than those of SZ3. Moreover, precision-progressive decompression does not reduce the size of the decompressed data, as the output retains its original resolution and data type (e.g., FP32 remains FP32). This still leads to significant I/O and memory overhead during decompression and post-processing. For these reasons, in this work, we focus on resolution-progressive decompression.

\textit{Resolution-progressive decompression} enables reconstructing a lower-resolution (coarse) version of the data and then progressively increasing the resolution. However, this approach incurs significant overhead in compression quality or speed. Resolution-progressive decompression can be achieved through wavelet transform, which de-correlates and decomposes the data into wavelet coefficients that can be progressively reconstructed from coarse to fine resolutions. For example, SPERR uses the CDF 9/7 wavelet transform, which provides both high compression quality and progressive capability. However, the wavelet transform in SPERR incurs high computational costs, resulting in significantly low compression throughput---up to 37$\times$ slower than our solution.

Another approach to resolution-progressive decompression is to directly partition the data into hierarchical levels.  For instance, AMM~\cite{bhatia2022amm} uses tree structures to partition the domain. However, such partitioning can break spatial locality and degrade compression quality and ratio (in lossy compression, compression quality and ratio are interchangeable—one can be traded off for the other). As a result, AMM cannot achieve compression quality comparable to SOTA compressors like ZFP and MGARD.
MGARD also adopts a hierarchical decomposition strategy but cannot eliminate the spatial locality loss due to partitioning, resulting in lower compression ratios than SZ3 and our solution. Moreover, MGARD has high computational complexity and is up to $18\times$ slower than our work.

Our solution also partitions the data hierarchically to enable high-speed resolution-progressive decompression.
However, to \textcolor{black}{achieve high compression quality},  we introduce an effective hierarchical prediction method to eliminate the negative impact of partitioning and achieve higher compression quality or ratio (e.g., up to 4.9$\times$ higher compression ratio than MGARD with the same data quality).

\subsection {Random Access Decompression}
\label{sec:back-rd}
Random access decompression enables independent decompression of specific regions of interest (ROI) in the dataset at full resolution, thereby reducing memory usage, decompression time, and I/O cost.
ZFP supports random-access decompression due to its block-wise design. Specifically, it partitions the data into $4^3$ blocks and performs transformation independently on each block, with no dependencies between them. However, this partitioning reduces spatial locality across blocks, leading to low compression quality.
In contrast, non-block-wise compressors such as SZ3, MGARD, SPERR, and TTHRESH preserve more spatial information but introduce strong data dependencies, making them incompatible with random-access decompression.

Our solution bridges this gap by enabling random-access decompression for non-block-wise compression that has high compression quality.
\textcolor{black}{That is, we achieve the non-block-wise compressor's high quality and the block-wise compressor's random-access decompression capability, with high performance, simultaneously.}
More specifically, our proposed hierarchical partitioning and prediction method introduces minimal dependency: all data points depend only on 1.6\% of the dataset. Thus, random-access decompression can be achieved by incurring only a 1.6\% prediction overhead.

For the encoding stage, we adopt Huffman encoding due to its high speed and high compression ratio. However, Huffman encoding does not support random-access decoding, as it requires decoding from the beginning of the encoded bitstream due to its variable-length nature.
One possible way to address this is to apply Huffman encoding to smaller chunks of data to enable independent decoding. However, this approach will significantly reduce compression ratios.
Thus, we choose to maximize the compression ratio by sacrificing little performance in random-access decoding.
Note that our solution still achieves up to 57\% savings in decoding time in many random-access decompression scenarios (i.e., accessing a 2D slice from a 3D dataset). Detailed analysis of random-access decompression savings is provided in \S\ref{sec:rd-acc} and \S\ref{sec:eva-str}.

\begin{figure}[h]
  \centering
  \includegraphics[width=\linewidth]{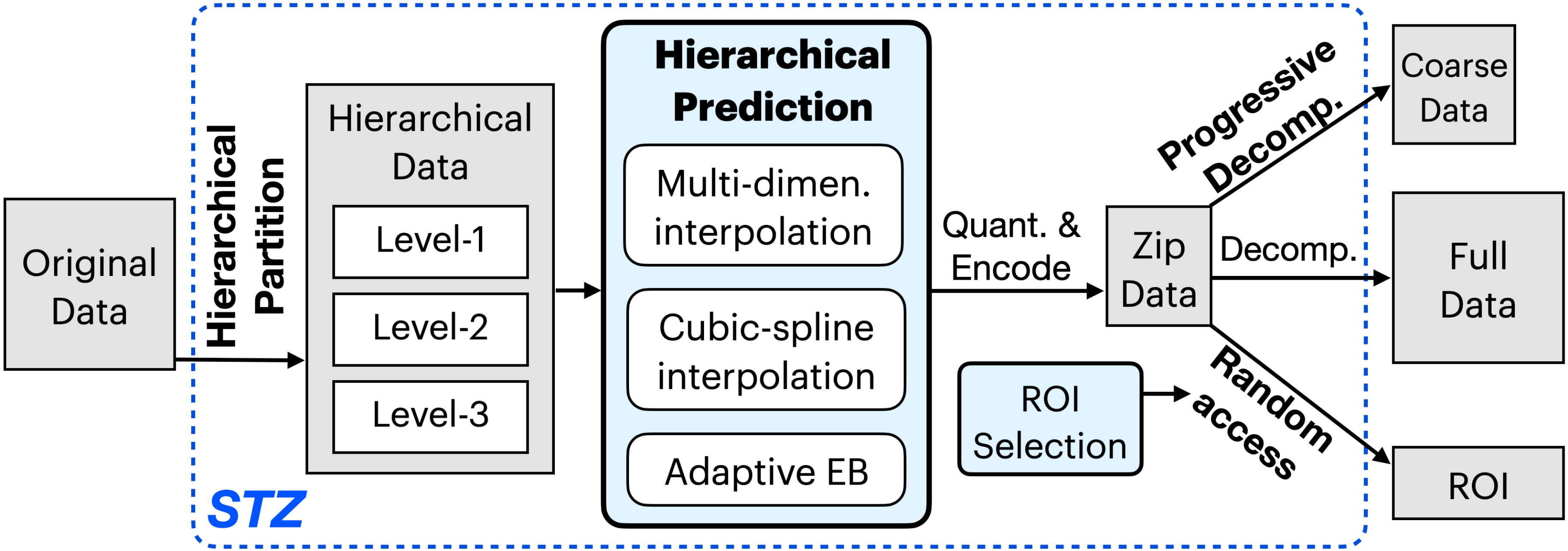}
  \caption{Overview of our proposed STZ}
  \label{fig:overview}
\end{figure}

\section{Design Methodology}

In this section, we introduce our proposed streaming lossy compression framework, \thiswork{}, illustrated in Figure~\ref{fig:overview}, which provides high compression quality and speed. The outline is detailed below.

In \S\ref{sec:pred-op}, we introduce our hierarchical partition approach, which enables streaming compression. We then present a series of hierarchical prediction optimizations to mitigate the spatial locality loss from partitioning, allowing our solution to achieve compression quality comparable to the SOTA non-streaming compressor SZ3.

In \S\ref{sec:3lvl}, we extend our approach to a 3-level hierarchical partition, which further improves progressive decompression and random-access decompression capabilities, while also increasing compression and decompression speed compared to the 2-level partition, achieving up to 6.7$\times$ speed up compared to SZ3.

In \S\ref{sec:rd-acc}, we detail our framework's progressive decompression and random-access decompression features, which enable users to progressively decompress a coarse (lower-resolution) version of the data and selectively decompress only the region of interest (ROI). We also propose an ROI selection module to identify the ROI.

\subsection{Hierarchical Partition \& Prediction}

\begin{figure*}[ht]
  \centering
  \begin{subfigure}[t]{0.24\linewidth}
    \centering
    \includegraphics[width=\linewidth]{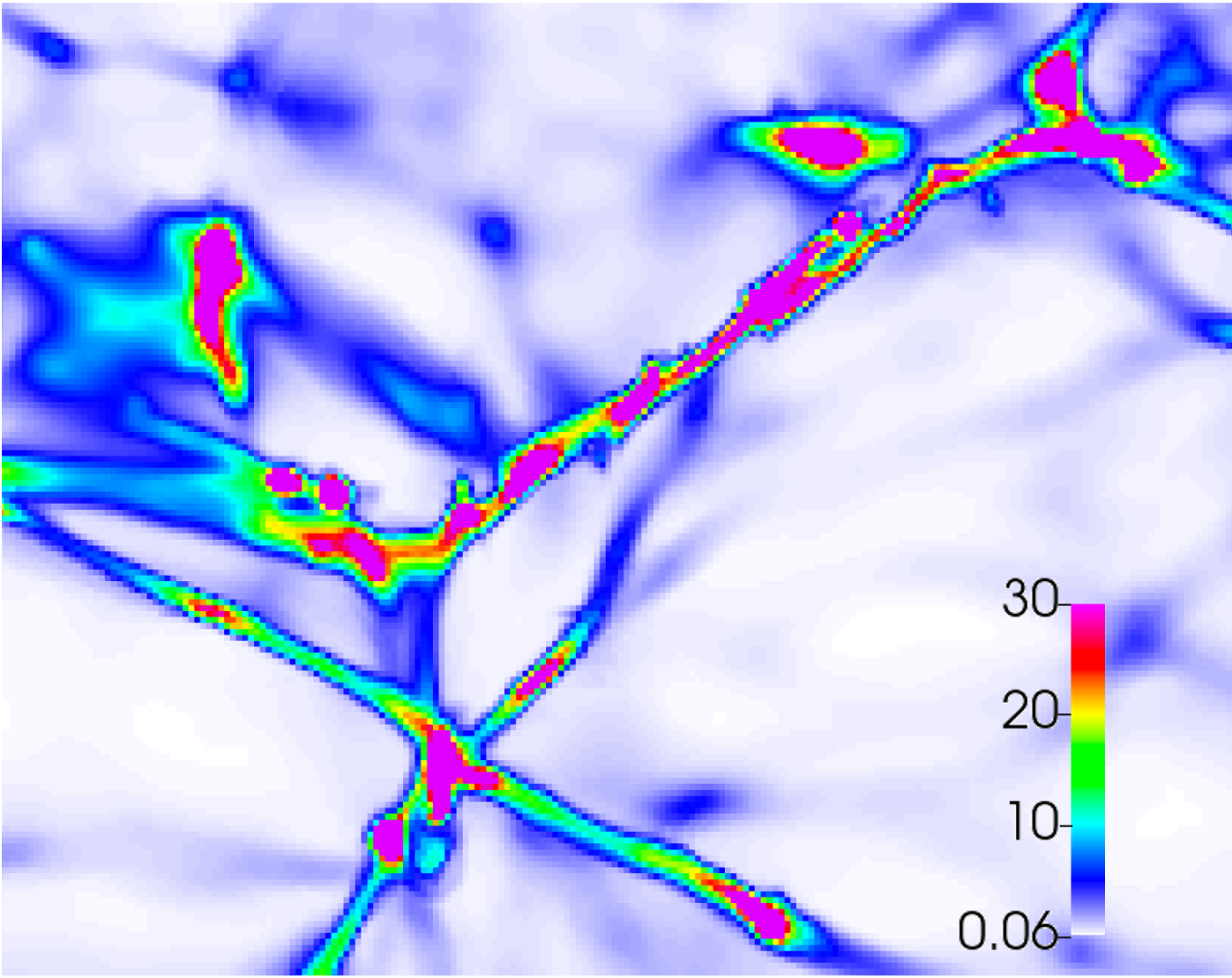}
    \caption[t]{Original data}
    \label{fig:vis-ori-slides}
  \end{subfigure}~%
  \begin{subfigure}[t]{0.24\linewidth}
    \centering
    \includegraphics[width=\linewidth]{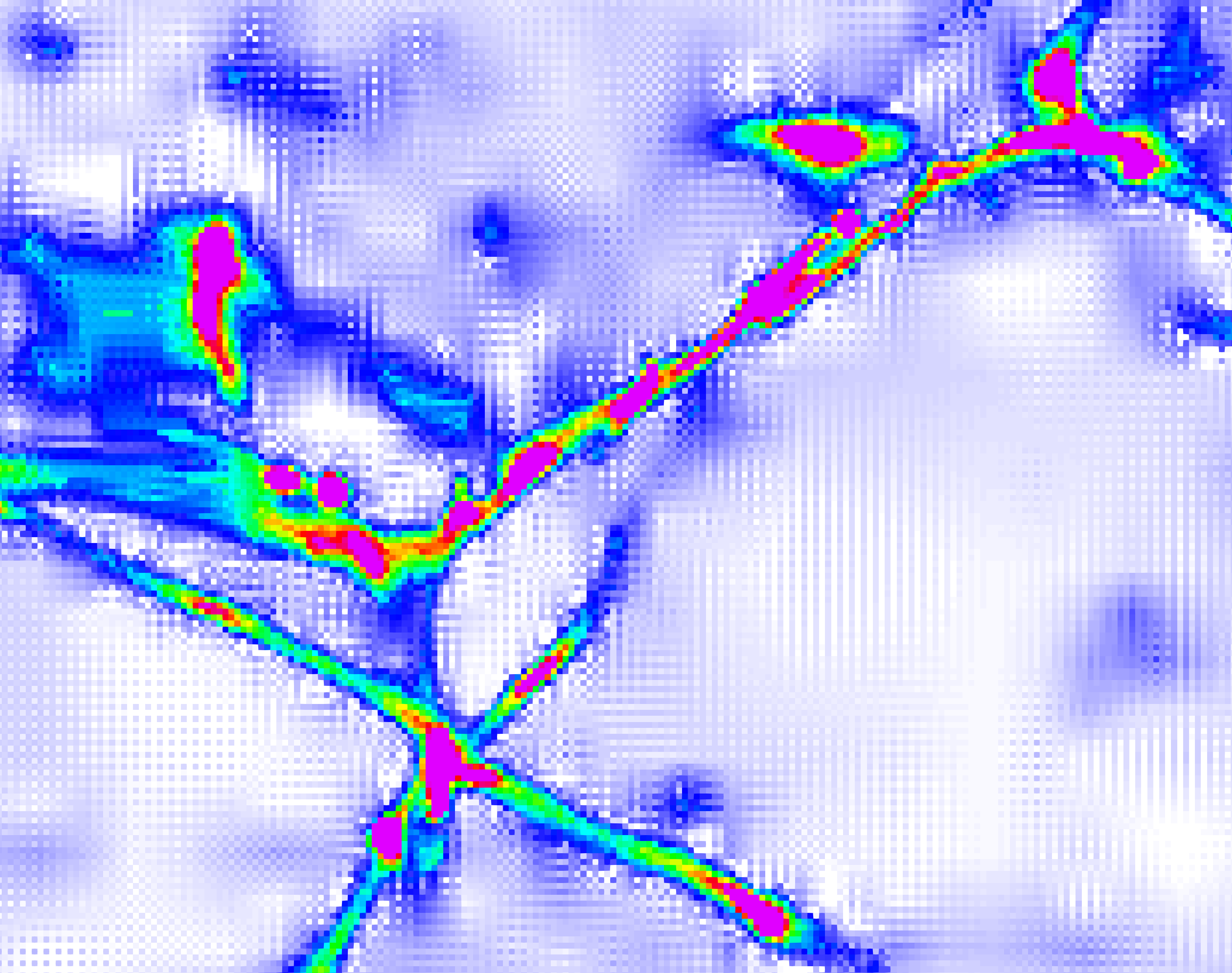}
    \caption{Partition, SSIM=0.67, PSNR=107}
    \label{fig:vis-parti-slides}
  \end{subfigure}~%
  \begin{subfigure}[t]{0.24\linewidth}
    \centering
    \includegraphics[width=\linewidth]{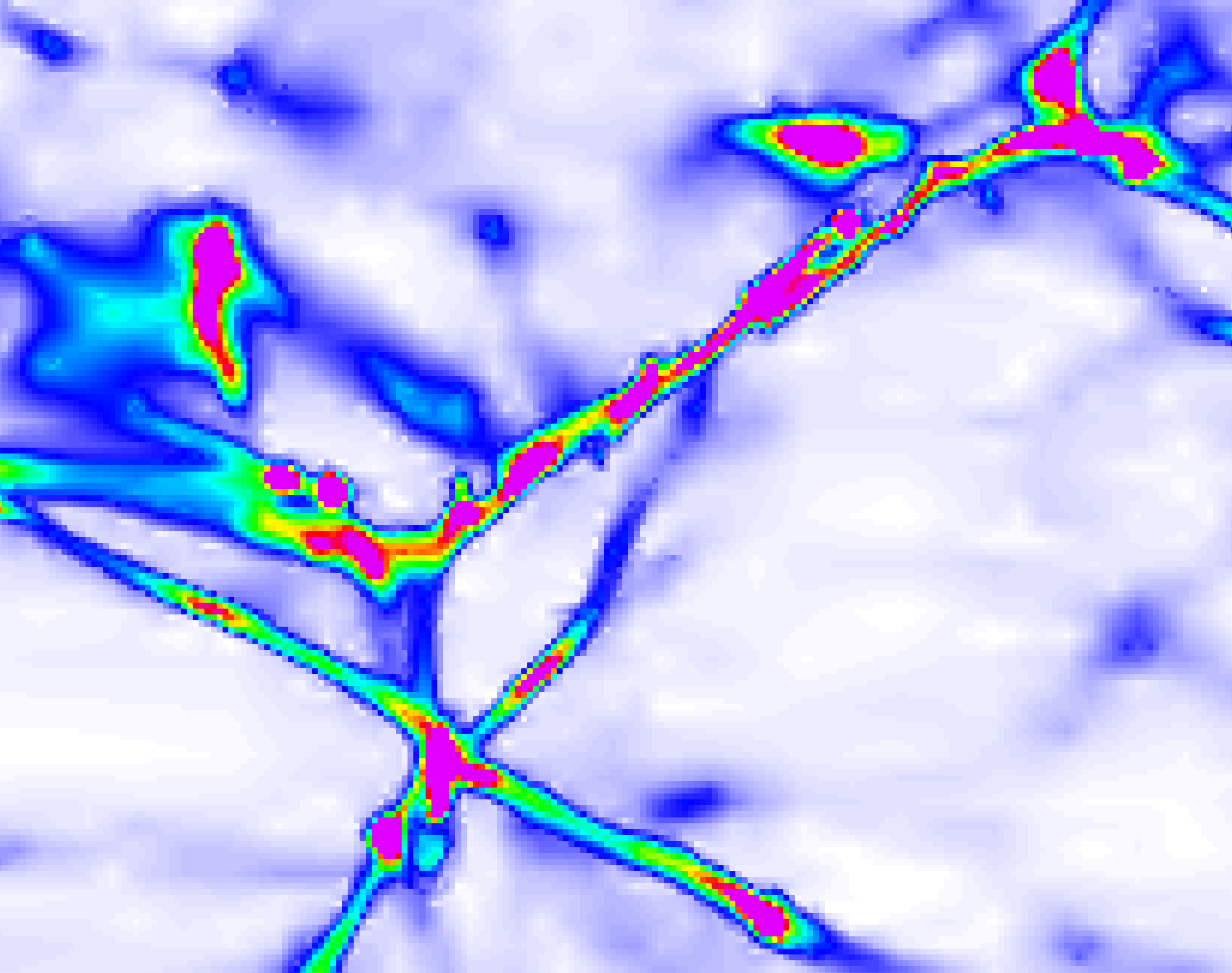}
    \caption{SZ3, SSIM=0.95, PSNR=118}
    \label{fig:vis-sz3-slides}
  \end{subfigure}~%
  \begin{subfigure}[t]{0.24\linewidth}
    \centering
    \includegraphics[width=\linewidth]{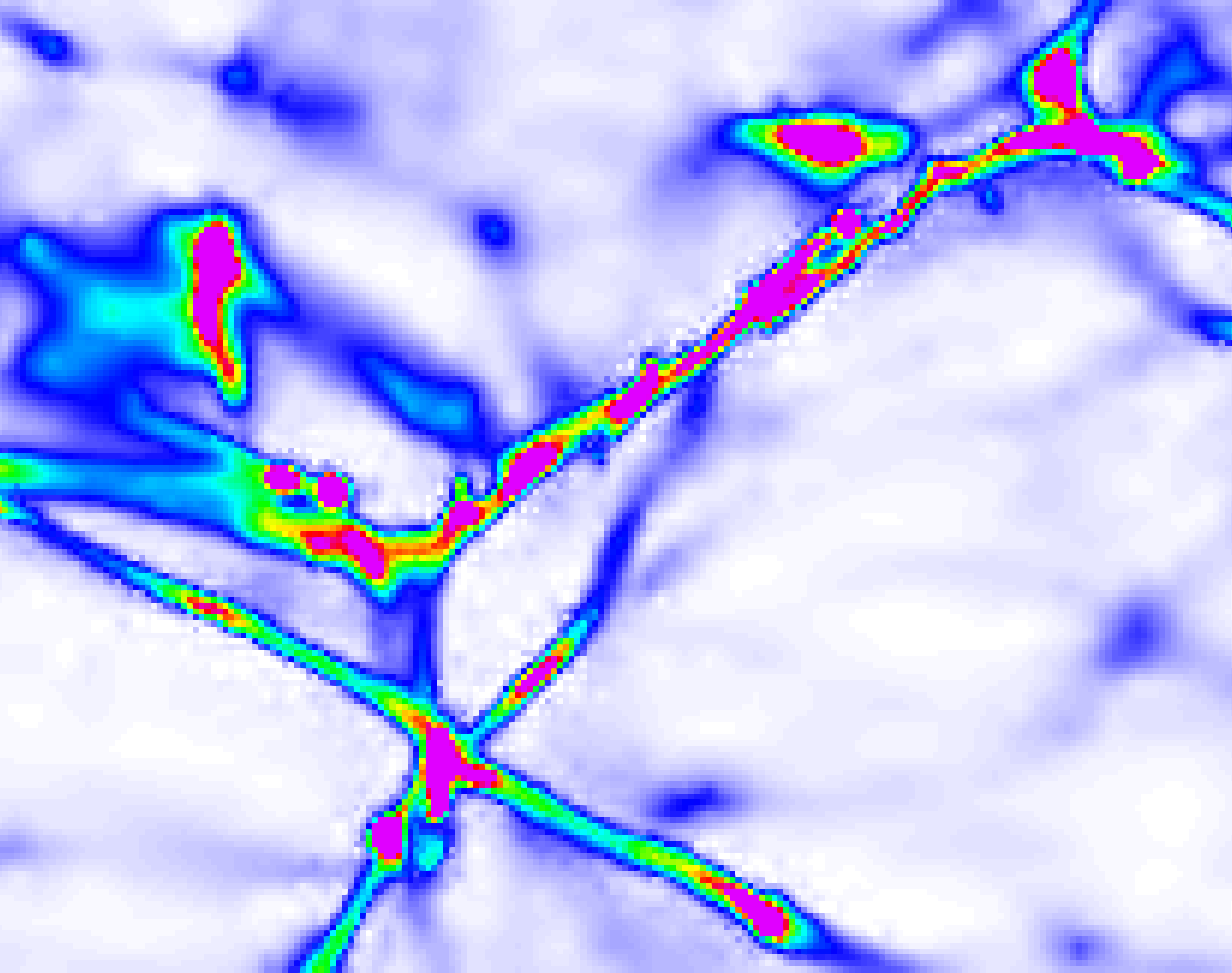}
    \caption{Ours, SSIM=0.95, PSNR=120}
    \label{fig:vis-ours-slides}
  \end{subfigure}
  \caption[t]{Visual comparison (one $2\times$ zoom in 2D slice) of original data and decompressed data produced by partition, SZ3, and our STZ on Nyx's ``baryon density'' field. The ``Rainbow Blended White'' colormap in ParaView is used. The CR of partition, SZ3, and ours are 204, 205, and 206.}
  \label{fig:vis-nyx-slides}
\end{figure*}

\label{sec:pred-op}
The progressive decompression feature allows users to decompress coarse representations of a dataset, reducing memory footprint and saving time during decompression and post-analysis when full-resolution recovery is unnecessary. To enable progressive decompression, data can be partitioned into multiple levels; however, this partitioning can disrupt spatial locality and reduce compression quality, as illustrated in Figure~\ref{fig:vis-parti-slides}; it can lead to significant distortion in the reconstructed data.
Besides partitioning, SPERR uses wavelet transforms to decompose data. While wavelet transforms can leverage more spatial information and enable progressive decompression, they are computationally expensive and result in very low compression speeds (up to 37$\times$ slower than our solution).

To achieve high speed, our solution adopts direct partitioning instead of using the wavelet transform. We then propose a hierarchical prediction approach with a series of optimizations to mitigate the loss of spatial locality caused by partitioning and to achieve significantly better compression quality, as shown in Figure~\ref{fig:vis-ours-slides}.
We now detail the challenges introduced by partitioning and introduce our optimization methods.

\begin{figure}[h]
  \centering
  \includegraphics[width=0.65\columnwidth]{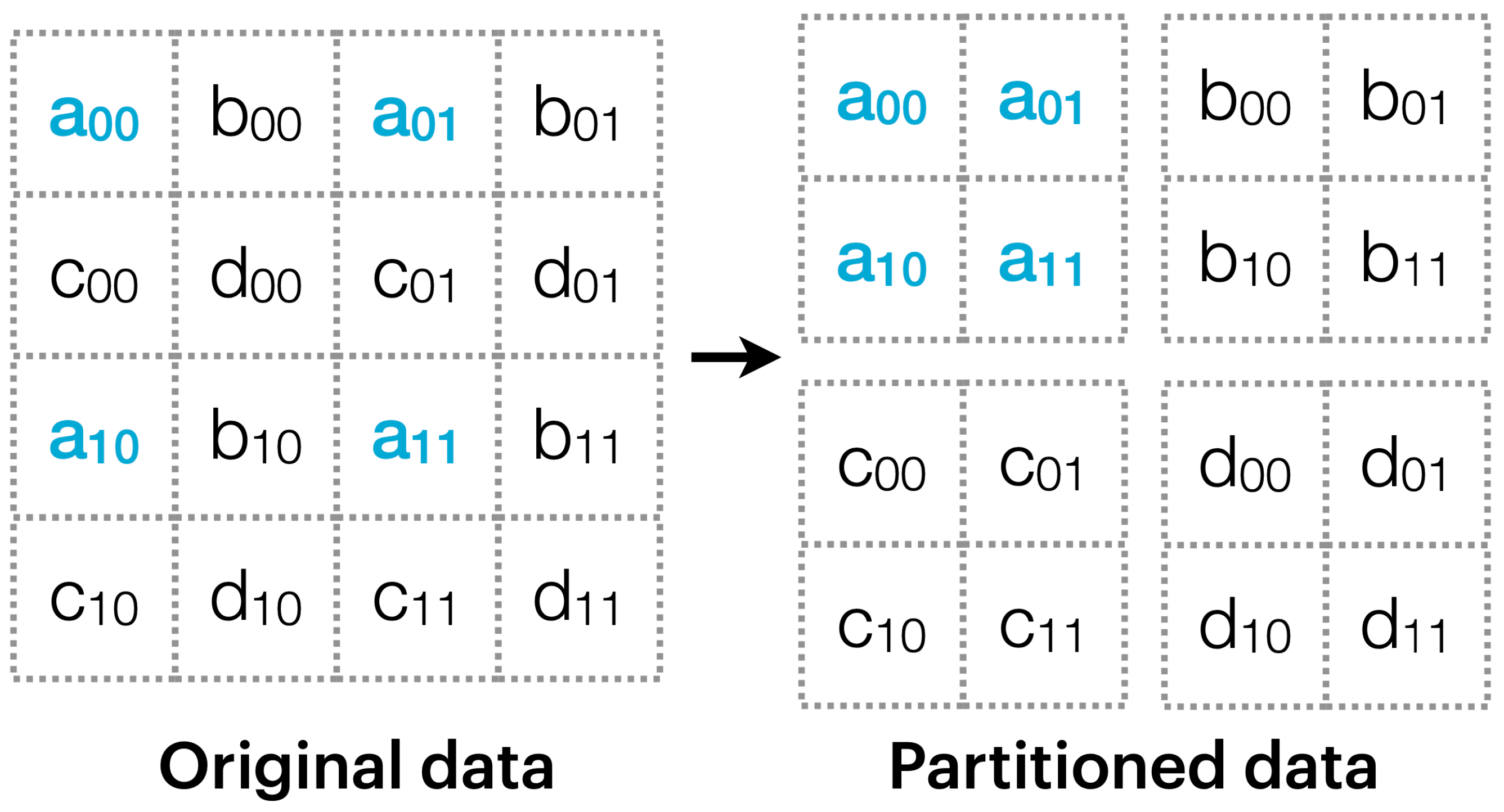}
  \caption{2D example of 2-level partition.}
  \label{fig:split}
\end{figure}

\textbf{Hierarchical partition and challenges.}
To enable efficient progressive decompression and random-access decompression, we avoid the slow wavelet transform used in SPERR. Instead, we start by directly partitioning the original 3D data into 8 smaller sub-blocks (4 blocks for 2D data), each having half the resolution along every dimension. As shown in Figure~\ref{fig:split}, the partition uses a stride-2 sampling along all axes with different starting offsets to ensure no overlap. After partition, each sub-block serves as a coarse representation of the original data. Progressive compression is achieved by compressing each sub-block separately using SZ3; decompressing just one sub-block (e.g., sub-block $a$) yields a coarse version of the data, while full-resolution data can be recovered by decompressing all the sub-blocks and reassembling them.

\begin{figure}[t]
  \includegraphics[width=\columnwidth]{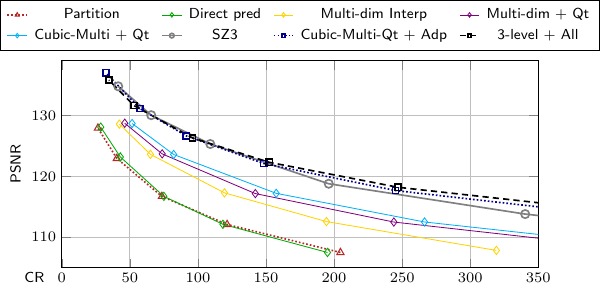}
  \caption{Rate-distortion comparison of direct partition, our optimizations, and SZ3 on the Nyx dataset (top-right is better).}
  \label{fig:rd}
\end{figure}

However, this partitioning process disrupts spatial information and leads to lower compression quality. As shown in Figure~\ref{fig:rd}, the overall rate-distortion performance of the naive partitioning method (labeled ``Partition'', red triangle) is significantly worse than the original SZ3 (i.e., using SZ3 to compress the unpartitioned data, labeled ``SZ3'', gray cycle). Figure~\ref{fig:vis-nyx-slides} further illustrates that the data quality produced by the partitioning method is greatly worse than that of SZ3. Additionally, compressing each sampled sub-block separately can result in inconsistent compression errors and introduce noticeable ``pixelated'' artifacts as shown in Figure~\ref{fig:vis-parti-slides}.

\textbf{Prediction Optimization 1: Direct Prediction.} To improve the quality of the decompressed data, we propose a method to better exploit the spatial information among partitioned sub-blocks. Each sub-block is highly similar to others because their sampling offsets differ by only 1 to 3 Manhattan units in the original grid (in the 3D case). As a result, redundancy exists among the sub-blocks, which can be leveraged to enhance compression quality. Specifically, we use one sub-block as a reference to predict the others and then compress the difference between the predicted and actual values.

\begin{figure}[h]
  \centering
  \includegraphics[width=0.63\columnwidth]{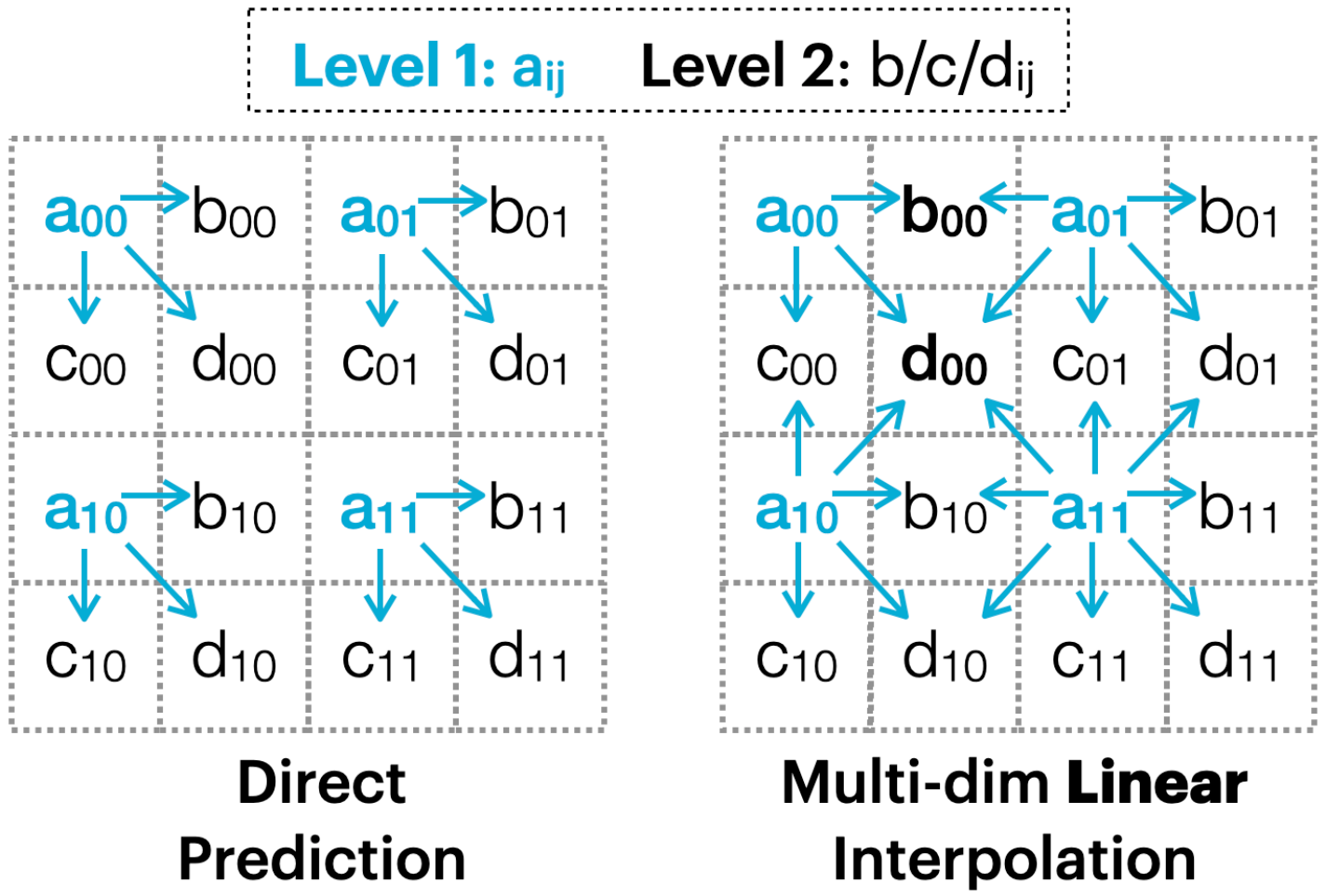}
  \caption{2D example of 2-level direct prediction (left) and multi-dimensional linear interpolation prediction (right).}
  \label{fig:pred-1}
\end{figure}

For example, as shown in the left part of Figure~\ref{fig:pred-1}, a straightforward approach is to use the first sub-block \(a\) as level 1 to directly predict the sub-blocks \(b/c/d\) at level 2 (i.e., points in level 2 are predicted using the corresponding point in sub-block \(a\)), such that:
\begin{equation}
  pred_{z,y,x} = a_{z,y,x}.
  \label{eq:1}
\end{equation}

In this case, the value to be compressed is the difference between the actual and predicted values. For instance, for sub-block $b$:
\begin{equation}
  diff_{z,y,x} = b_{z,y,x} - pred_{z,y,x} = b_{z,y,x} - a_{z,y,x}.
  \label{eq:2}
\end{equation}
During decompression, $b_{z,y,x}$ is recovered using $a_{z,y,x}$ and the difference $diff_{z,y,x}$.

Although the similarity between sub-blocks results in relatively small and smooth differences, which improve compression effectiveness, the rate-distortion of the direct prediction method (labeled ``Direct pred'', green diamond), as shown in Figure~\ref{fig:rd}, offers only little improvement over the partition method, and only at low compression ratios (CR). Consequently, the compression performance of the direct prediction approach remains significantly lower than that of the original SZ3.

\textbf{Prediction Optimization 2: Multi-dimensional Interpolation.} This approach's main motivation is to leverage additional unused spatial information. As shown in the right part of Figure~\ref{fig:pred-1}, the data point $b_{00}$ lies in middle between $a_{00}$ and $a_{01}$. Therefore, instead of relying only on $a_{00}$ to directly predict $b_{00}$, we use both $a_{00}$ and $a_{01}$ via linear interpolation to leverage more spatial information and improve compression as follows:
\begin{equation}
  \text{pred}_{z,y,x} = \frac{1}{2}a_{z,y,x} + \frac{1}{2}a_{z,y,x+1}
  \label{eq:3}
\end{equation}

For the data point $d_{00}$, which lies in the middle between $a_{00}$, $a_{01}$, $a_{10}$, and $a_{11}$, we use its four neighboring points to predict $d_{00}$ via bi-linear interpolation:
\begin{equation}
  \text{pred}_{z,y,x} = \frac{1}{4}a_{z,y,x} + \frac{1}{4}a_{z,y,x+1} + \frac{1}{4}a_{z,y+1,x} + \frac{1}{4}a_{z,y+1,x+1}
  \label{eq:4}
\end{equation}
The linear and bi-linear interpolation is applied to other points in level 2, except for the boundary points, which are predicted directly from available data (e.g., $d_{11}$ from $a_{11}$).

\begin{figure}[h]
  \vspace{-1mm}
  \centering
  \begin{subfigure}[t]{0.32\linewidth}
    \centering
    \includegraphics[width=\linewidth]{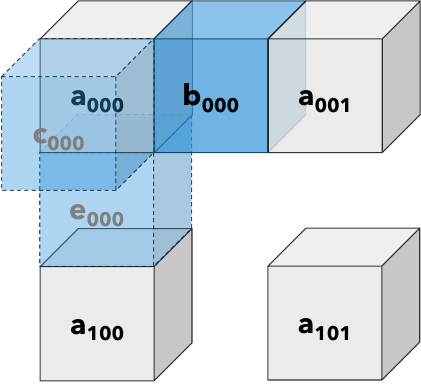}
    \caption[t]{Linear interpolation}
    \label{fig:linear}
  \end{subfigure}
  \begin{subfigure}[t]{0.32\linewidth}
    \centering
    \includegraphics[width=\linewidth]{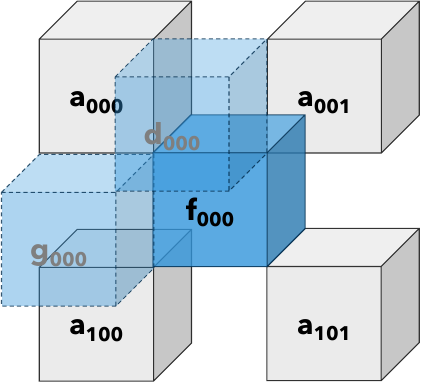}
    \caption{Bi-linear interp.}
    \label{fig:bi-linear}
  \end{subfigure}
  \begin{subfigure}[t]{0.34\linewidth}
    \centering
    \includegraphics[width=\linewidth]{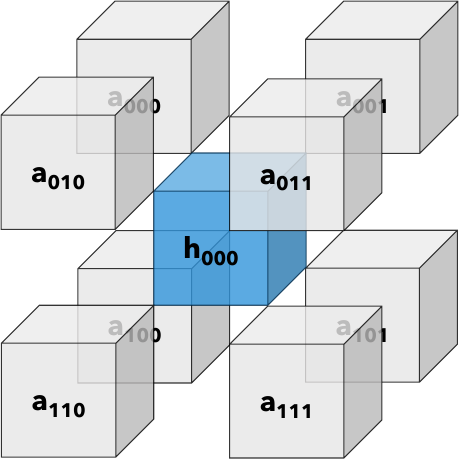}
    \caption{Tri-linear interpolation}
    \label{fig:tri-linear}
  \end{subfigure}
  \caption{Visualization of different interpolation cases in 3D data.}
  \label{fig:3d-interp}
\end{figure}

In the 3D case, there are seven sub-blocks at level 2 (the full-resolution level), as shown by the seven blue boxes in Figure~\ref{fig:3d-interp}.
Three out of the seven sub-blocks at level 2 (i.e., sub-blocks $b/c/e$) are positioned with a 1-Manhattan-unit offset along one of the X, Y, or Z dimensions relative to sub-block $a$, as shown in Figure~\ref{fig:linear}. The points in these sub-blocks are next to two data points in sub-block $a$. For example, $b_{000}$ is adjacent to $a_{000}$ and $a_{001}$.
Points in sub-blocks $b$ (and similarly in $c$ and $e$) are predicted using linear interpolation in Equation~\ref{eq:3} with the two neighboring data in sub-block $a$.

Another three sub-blocks ($d/f/g$) are positioned with a 2 Manhattan unit offset along two of the X, Y, or Z dimensions relative to sub-block $a$, as shown in Figure~\ref{fig:bi-linear}. The points in these sub-blocks are adjacent to four data points in $a$. For example, $f_{000}$ is adjacent to $a_{000}$, $a_{001}$, $a_{100}$, and $a_{101}$.
Points in sub-blocks $f$ (and similarly in $d$ and $g$) are predicted using bilinear interpolation, as defined in Equation~\ref{eq:4}, based on the four adjacent data points in sub-block $a$.

The final sub-block $h$ is positioned with a 3-Manhattan-unit offset along all three dimensions and is located diagonally to sub-block $a$, as shown in Figure~\ref{fig:tri-linear}. Each point in this sub-block is diagonally adjacent to eight data points in sub-block $a$ and is tri-linearly interpolated using the eight adjacent points:
\begin{equation}
  \text{pred}_{x,y,z} = \frac{1}{8} \sum_{k=0}^{1}\sum_{j=0}^{1}\sum_{i=0}^{1} a_{z+k,\,y+j,\,x+i}
  \label{eq:5}
\end{equation}

As shown in Figure~\ref{fig:rd}, multi-dimensional linear interpolation (labeled ``Multi-dim Interp'', yellow diamond) can better leverage the spatial redundancy between level 1 and level 2 and improve the rate-distortion of the direct prediction.

\textbf{Prediction Optimization 3: Remove the SZ3 Prediction.} We further propose removing SZ3’s original prediction step for level 2 (i.e., sub-blocks $b/c/d$ in the 2D examples). SZ3 consists of three steps: first, it uses 1D spline interpolation along each dimension to predict the data; next, it quantizes the difference between the predicted and actual values; finally, it applies Huffman encoding to these differences.

For the sub-blocks at level 2, predictions have already been made using sub-block $a$, and their spatial redundancy has been exploited. Therefore, there is no need to perform SZ3’s prediction step again. Specifically, after prediction using sub-block $a$, the difference between the predicted and original values at level 2 is saved for compression. These differences are relatively “random” (i.e., they exhibit little correlation). In such cases, performing a second round of SZ3 prediction would (1) waste computational time and (2) degrade compression quality. As shown in Figure~\ref{fig:rd} (labeled ``Multi-dim + Qt'', purple diamond), removing SZ3’s original prediction and applying only quantization and encoding during compression further improves the performance of our method.

\textbf{Prediction Optimization 4: Cubic Interpolation.} We propose using cubic spline interpolation to further improve the prediction of linear interpolation. Cubic spline interpolation leverages more spatial information than linear interpolation and yields better performance when estimating a new data point using existing points.

\begin{figure}[h]
  \vspace{-1mm}
  \centering
  \includegraphics[width=0.75\columnwidth]{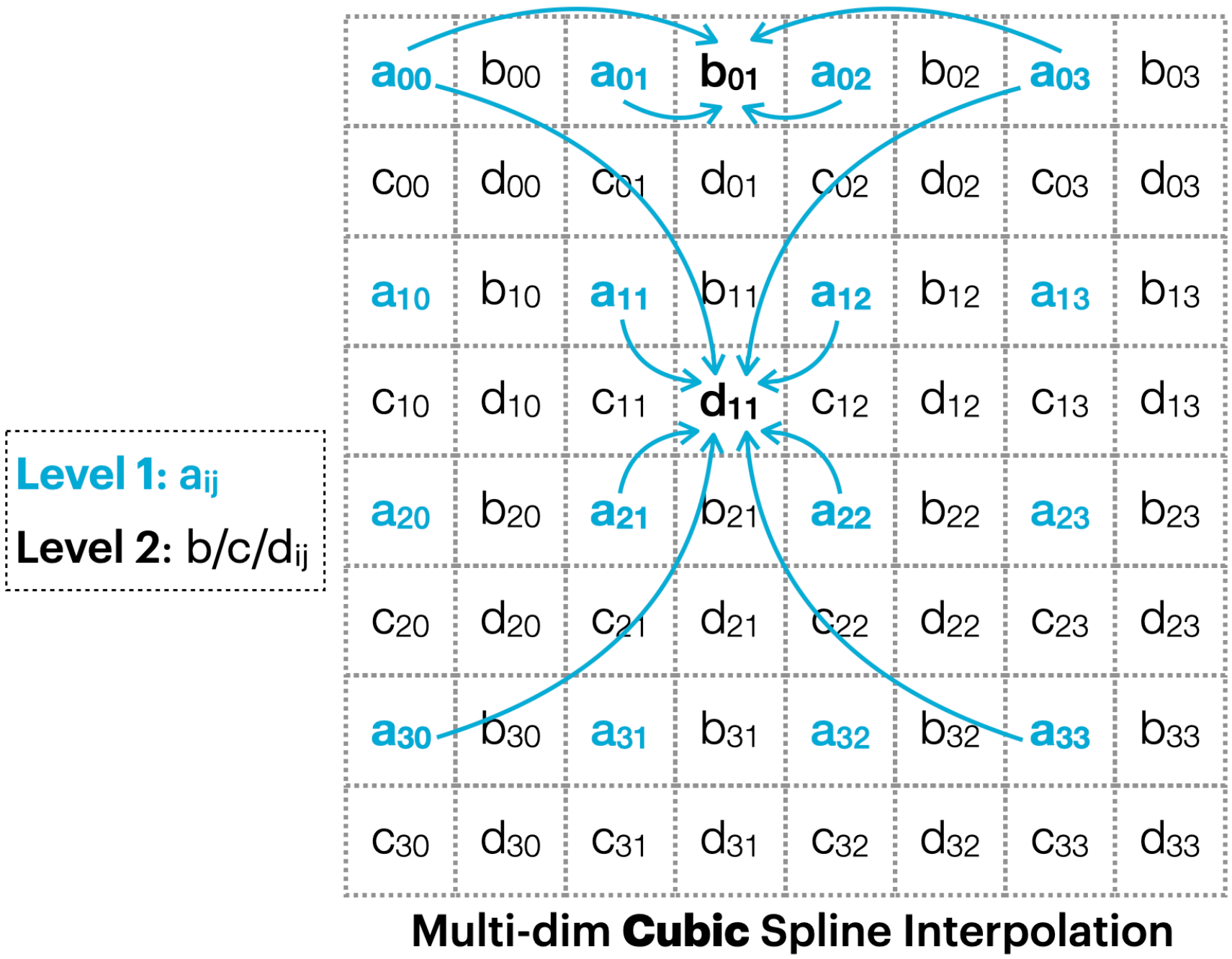}
  \caption{2D example of multi-dimensional cubic spline interpolation prediction for $b_{01}$ and $d_{11}$, not all interpolations are shown for clarity.}
  \label{fig:pred-2}
  \vspace{-1mm}
\end{figure}

For example, as shown in Figure~\ref{fig:pred-2}, to predict the data point $b_{01}$, in addition to $a_{01}$ and $a_{02}$, we can also use $a_{00}$ and $a_{03}$ from 3 units away and apply cubic spline interpolation as follows:
\begin{equation}
  \text{pred}_{z,y,x} = -\frac{1}{16}a_{z,y,x-1} + \frac{9}{16}a_{z,y,x} + \frac{9}{16}a_{z,y,x+1} - \frac{1}{16}a_{z,y,x+2}
  \label{eq:6}
\end{equation}
We refer the reader to~\cite{zhao2021optimizing} for the detailed derivation of the cubic spline interpolation formula.
In the 3D case, as shown in Figure~\ref{fig:linear}, sub-block $b$ is also predicted via cubic spline interpolation equation~\ref{eq:6}. Sub-blocks $c$ and $e$ follow a similar approach but apply the interpolation along different dimensions.

For data point $d_{11}$, we apply an approximate bi-cubic spline interpolation using the data points $a_{11}$, $a_{21}$, $a_{12}$, $a_{22}$, $a_{00}$, $a_{30}$, $a_{03}$, and $a_{33}$, such that:
\begin{equation}
  \text{pred}_{x,y,z} =
  \frac{9}{32} \sum_{j,i=0}^{1} a_{y+j,\,x+i,\,z}
  -
  \frac{1}{32} \sum_{j,i \in \{-1,2\}} a_{y+j,\,x+i,\,z}
  \label{eq:7}
\end{equation}
This is achieved by combining two cubic spline interpolations: one along the diagonal formed by $a_{00}$, $a_{11}$, $a_{22}$, and $a_{33}$, and another along the orthogonal diagonal formed by $a_{03}$, $a_{12}$, $a_{21}$, and $a_{30}$. Each cubic spline interpolation contributes half of the weight. In the 3D case, as shown in Figure~\ref{fig:bi-linear}, sub-block $f$ is also predicted via cubic spline interpolation using equation~\ref{eq:7}, with sub-blocks $d$ and $g$ following the same approach along different dimensions.

\begin{figure*}[t]
  \centering
  \includegraphics[width=0.88\linewidth]{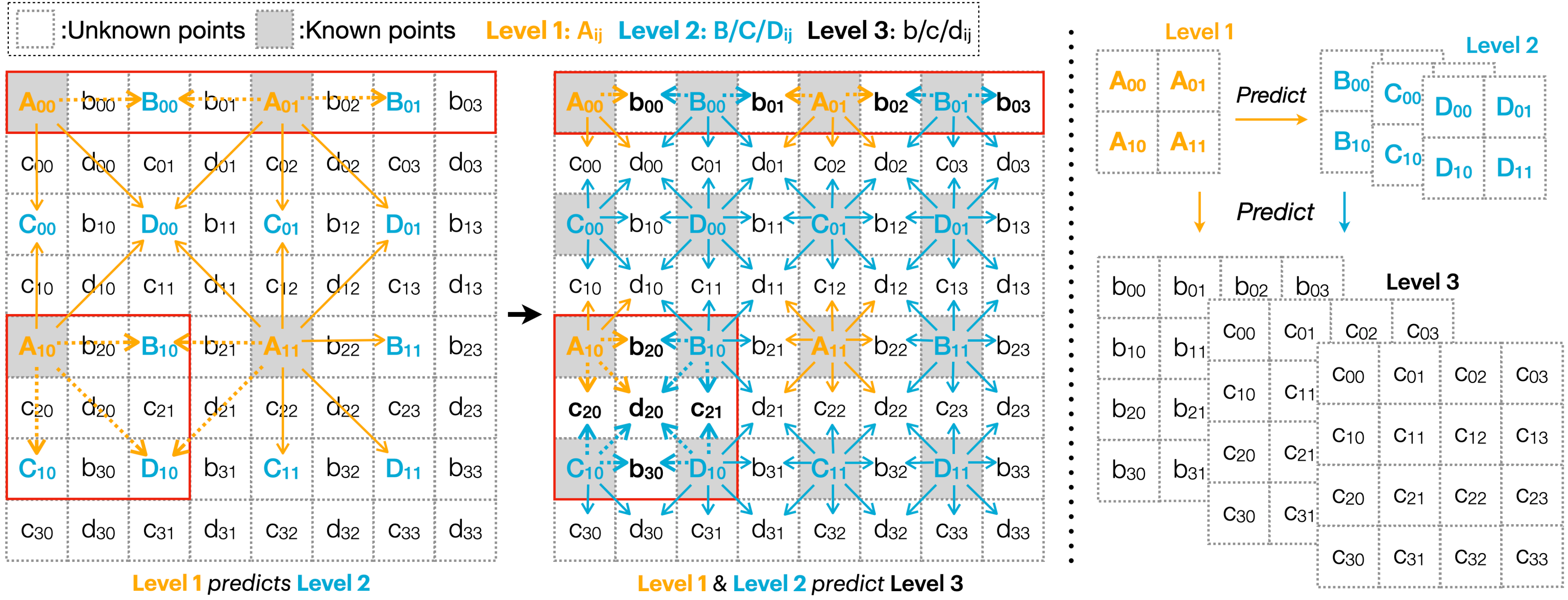}
  \caption{2D example of three-level partition (right) and hierarchical interpolation prediction (left~\&~center), linear interpolation was used for clarity. Level 1 (orange) predicts level 2 (blue), then levels 1 and 2 predict level 3.
  Red boxes and dashed arrows highlight examples of ROIs that can be decompressed independently via random-access decompression (i.e., only the predictions indicated by the dashed arrows are required).}
  \label{fig:3lvl}
  \vspace{-1mm}
\end{figure*}

In the 3D case, where sub-block $h$ is positioned diagonally relative to $a$, as shown in Figure~\ref{fig:tri-linear}.
As mentioned, Points in sub-block $h$ are adjacent to eight data points in $a$ and can be tri-linearly interpolated using these eight points.
To further apply an approximate tri-cubic spline interpolation, besides the eight neighboring points, eight additional outer-layer points are used for prediction (i.e., $\sum_{k,j,i \in \{-1,2\}}
a_{z+k,\,y+j,\,x+i}$).
This forms a combination of four cubic spline interpolations, each contributing 25\% of the total weight:
{\normalsize
  \begin{equation}
    \text{pred}_{x,y,z} =
    \frac{9}{64} \sum_{k,j,i=0}^{1}
    a_{z+k,\,y+j,\,x+i}
    -
    \frac{1}{64} \sum_{k,j,i \in \{-1,2\}}
    a_{z+k,\,y+j,\,x+i}
    \label{eq:8}
  \end{equation}
}
As shown in Figure~\ref{fig:rd}, cubic spline interpolation (labeled ``Cubic-Multi + Qt'', light blue diamond) can further improve the compression performance compared to linear interpolation.

\textbf{Prediction Optimization 5: Adaptive Error Bound.} The final optimization we propose is the use of adaptive error bounds across different hierarchical levels. Since level 1 (sub-block $a$) is used to predict level 2 (sub-blocks $b/c/d$ in 2D cases), its compression error propagates to level 2's prediction. Thus, using a smaller error bound to level 1 can improve overall compression efficiency. Additionally, preserving high data quality in level 1 is a worthwhile trade-off, as it is frequently accessed in streaming decompression due to its role as a coarse representation that can be directly decompressed.

To determine optimal error bounds for levels 1 and 2, we perform extensive experiments on different datasets and test various error bound ratios between the two levels. Our results show that when the error bound for level 2, \(eb_{l2}\), is set to 2.5 times the error bound for level 1, \(eb_{l1}\), i.e., \(eb_{l2} = 2.5 \times eb_{l1}\), the overall best compression performance is achieved. As shown in Figure~\ref{fig:rd} (labeled ``Cubic-Multi-Qt + Adp'', deep blue dotted square), the adaptive error bound allows our solution to achieve performance similar to SZ3 while yielding better data quality at higher compression ratios. As shown in Figure~\ref{fig:vis-ours-slides}, after applying all five optimizations, our solution closes the data quality gap between the direct partition method and the original SZ3, while achieving streaming compression capability.

\subsection{Three-level Partition \& Prediction}
\label{sec:3lvl}
The previously described partition and prediction method in \S\ref{sec:pred-op} (i.e., Figures~\ref{fig:split},~\ref{fig:pred-1}, and~\ref{fig:pred-2}) uses a two-level scheme. Specifically, sub-block $a$ represents the coarse representation (level 1) and is used to predict the other sub-blocks ($b$, $c$, and $d$) at level 2.

However, the coarsest level 1 still takes 12.5\% of all the 3D data, meaning users must decompress at least 12.5\% of the data to get a coarse representation, even when less data is sufficient, limiting the streaming capability.
To further improve the flexibility of progressive decompression, allowing data to be decompressed at even lower resolutions, and to enhance both compression and decompression speed, we propose a three-level partitioning and prediction approach, as shown in Figure~\ref{fig:3lvl}.

In the three-level approach, the entire dataset (using 2D data as an example) is partitioned into seven sub-blocks: $A$, $B$, $C$, $D$, $b$, $c$, and $d$. Sub-block $A$ serves as the coarsest level 1, obtained using stride-4 sampling (the coarsest representation, marked in gray in the left of Figure~\ref{fig:3lvl}), and is used to predict sub-blocks $B$, $C$, and $D$ at level 2. Sub-blocks $A$, $B$, $C$, and $D$ together form a coarse representation using stride-2 sampling (marked in gray in the center of Figure~\ref{fig:3lvl}), which is then used recursively to predict sub-blocks $b$, $c$, and $d$ at level 3.
The prediction mechanism between each level in the three-level method remains the same as in the two-level case. It uses multi-dimensional cubic spline interpolation combined with quantization and adaptive error bounds to ensure high compression quality.
As shown in Figure~\ref{fig:rd}, the 3-level partition method (labeled ``3-level + All'', black dashed square) achieves compression quality similar to the 2-level partition and the original SZ3, with slightly better peak signal-to-noise ratio (PSNR) when the compression ratio is relatively high.

Furthermore, the compression and decompression speed of the 3-level partition is up to 2.2$\times$ faster than that of the 2-level partition.
This improvement is due to the speed difference between SZ3's and our prediction. In our framework, SZ3 is applied to the coarsest level 1, while our multi-dimensional prediction is used for the rest of the data. For a 2-level partition, SZ3 handles 12.5\% of the data, and for a 3-level partition, it handles only 1.6\%. Since our solution is faster than SZ3, applying it to a larger portion of the data results in an overall speedup for the 3-level partition. The speed of our solution and SZ3 will be detailed in \S\ref{sec:eva-speed}.

Our method supports more than three levels; however, for the tested dataset sizes, we do not propose a 4-level partition method, as a 3-level design already provides sufficient progressive decompression capability. For the 3-level design, the coarsest level accounts for only 1.6\% (1/64) of the dataset, leaving 98.4\% untouched during decompression. A 4-level design would increase this to 99.8\% (511/512), offering only marginal benefits. For much larger datasets (e.g., $4096^3$), we plan to offer an optional 4-level (or higher) version.

\subsection[Random Access \& Progressive Decompression]{Random Access De.~\& Progressive De.}
\label{sec:rd-acc}

\textbf{Progressive decompression. }
The hierarchical nature of our solution aligns well with the progressive decompression feature.
In Figure~\ref{fig:3lvl}, we decompose the original data into three levels with different resolutions. During decompression, users can choose to reconstruct only the lowest resolution by decompressing level 1. Users can then progressively recover higher resolutions by continuing to decompress level 2. The full-resolution data is obtained by additionally decompressing level 3 and assembling all three levels.

\textbf{Random access decompression.} Random access decompression helps users reduce memory usage, decompression time, and I/O costs by allowing only the required data to be decompressed. Block-wise compressors like ZFP support random-access decompression because they partition the data into blocks, and each block is processed independently, with no dependencies between blocks during decompression. However, this block-wise approach breaks the spatial locality of the dataset and significantly degrades compression quality. In contrast, non-block-wise compressors like SZ3, MGARD, and SPERR capture more spatial information across the dataset and achieve higher compression quality. However, due to the strong data dependencies introduced by their transformation and prediction processes, they are not compatible with random-access decompression.

Our solution overcomes this trade-off by enabling low-overhead random-access decompression for non-block-wise compression. This is achieved through our hierarchical partitioning and prediction. As discussed in \S\ref{sec:3lvl} and illustrated in Figure~\ref{fig:3lvl}, all data at levels 2 and 3 depend only on nearby data from the previous resolution level, with no prediction dependencies within the same level.

Specifically, after decompressing the coarsest level 1 (sub-blocks $A$), we can use it to selectively \textit{predict} only the ROI when reconstructing the full-resolution data. For example, as shown in Figure~\ref{fig:3lvl}, to decompress only the ROI indicated by the $3\times3$ red square (from $A_{10}$ to $D_{10}$), we first decompress level 1 and then use it to predict only $B_{10}$, $C_{10}$, and $D_{10}$ in level 2. Next, level 1 and the points just decompressed in level 2 will predict only $b_{20}$, $c_{20}$, $d_{20}$, $c_{21}$, and $b_{30}$ in level 3, skipping the prediction of all other points in levels 2 and 3. Although decompressing level 1 introduces overhead, level 1 takes only 1.6\% of the 3D data. In other words, our solution reduces the prediction overhead during decompression by up to 98.4\%.

As mentioned in \S\ref{sec:back-rd}, although we can achieve 98.4\% prediction savings in random-access decompression, \textit{decoding} time savings cannot always be achieved because, to achieve high compression ratios, all points in each sub-block are encoded together. Specifically, during compression,
levels 1 and 2 are used to predict level 3 (sub-blocks $b/c/d$), and their prediction errors are encoded. For high encoding efficiency, the prediction errors in each sub-block are encoded collectively. Therefore, during decompression to the red square from $A_{10}$ to $D_{10}$, to decode $b_{20}$, $c_{20}$, $d_{20}$, $c_{21}$, and $b_{30}$, we must decode all points in sub-blocks $b/c/d$.

On the other hand, decoding time can be saved when accessing a 1D ROI array, such as the $8\times1$ array from $A_{00}$ to $b_{03}$ in Figure~\ref{fig:3lvl}. In this case, for level 3, only $b_{00}$, $b_{01}$, $b_{02}$, and $b_{03}$ need to be decompressed. This means that only sub-block $b$ in level 3 needs to be decoded, as no points in sub-blocks $c$ or $d$ are involved. This results in a $\frac{2}{3}$ reduction in decoding time at level 3.
Note that in 3D datasets, the equivalent of that 1D array is a 2D slice, which is commonly accessed for visualization and post-analysis. Therefore, for 2D slice random-access decompression, our solution can achieve up to 98.4\% prediction savings and 58\% decoding savings. The detailed saving for random-access decompression will be presented in \S\ref{sec:eva-str}.

\begin{figure}[h]
  \centering
  \begin{subfigure}[t]{0.4\linewidth}
    \centering
    \includegraphics[width=\linewidth]{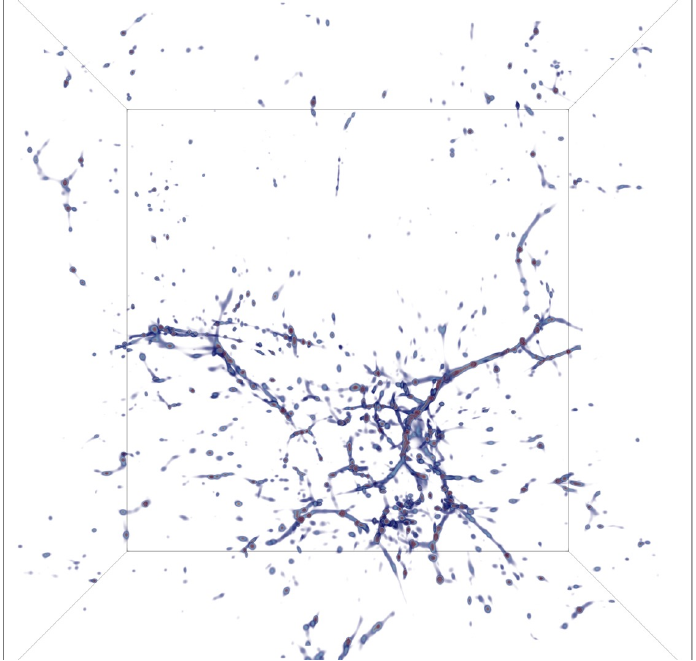}
    \caption[t]{Original data}
    \label{fig:roi-all}
  \end{subfigure}
  \begin{subfigure}[t]{0.4\linewidth}
    \centering
    \includegraphics[width=\linewidth]{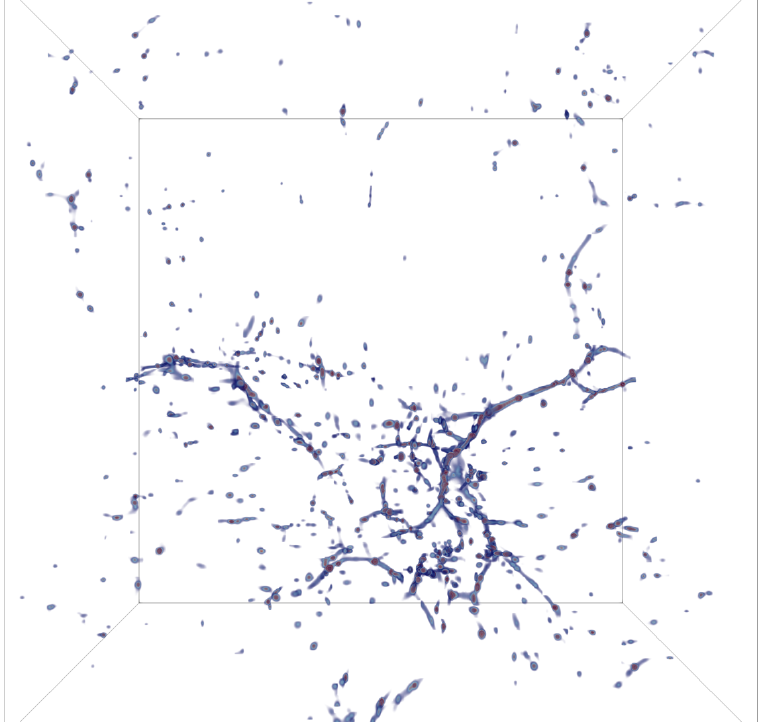}
    \caption{ROI, 0.69\%}
    \label{fig:roi-fine}
  \end{subfigure}

  \caption[t]
  {Vis of the ROI subset (right, 0.69\% of the dataset) extracted using our ROI module, compared to the full Nyx dataset (left).}
  \label{fig:roi}
\end{figure}

\begin{figure*}[t]
  \includegraphics[width=0.95\textwidth]{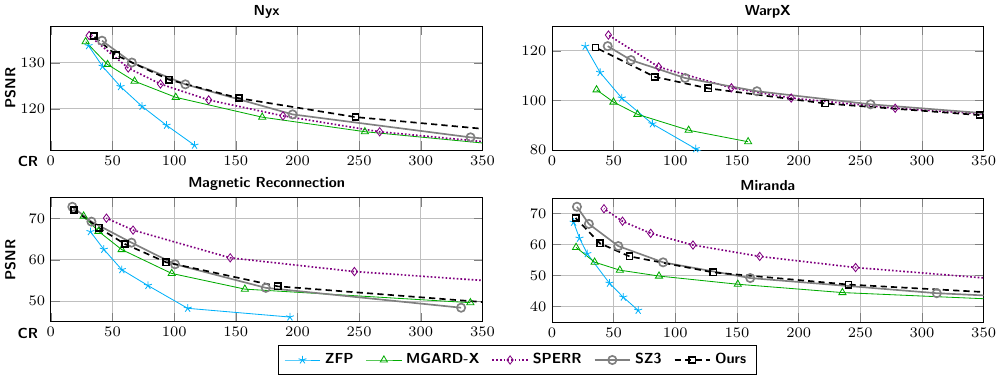}
  \caption[t]{Rate-distortion of our solution and baselines on different datasets (top-right is better). \textcolor{black}{Note that SPERR achieves high compression ratios on Magnetic Reconnection and Miranda, but its speed is significantly lower than other compressors, except for MGARD-X.}}
  \label{fig:all-rd}
\end{figure*}

\textbf{Flexible scientific workflow.}
The progressive decompression feature can also work with random-access decompression and enable a flexible scientific workflow when determining the ROI of the dataset. For example, by using our streaming compression, the user can first progressively decompress the coarsest version of the dataset for visualization and identify the ROI. After that, users can selectively decompress only ROI with full resolution. The coarsest data dumped by our solution only takes 1.6\% of the full data size, which can greatly save the decompression and visualization cost as well as the memory footprint.
This is especially important for users analyzing large-scale datasets on local computers with limited resources and for even larger datasets (e.g., petabyte-scale) on HPC systems, where decompressed data may exceed the memory capacity of multiple nodes.

Our framework also includes a module to help users to identify ROIs, such as 2D slices or 3D blocks.
Users can choose range-based or maximum-value thresholding, depending on the simulation.
For example, range thresholding suits fluid dynamic simulation because it effectively identifies the interface between two fluids of different densities, while maximum-value thresholding is more suitable for identifying the over-density halos in cosmology simulation.
We compute the value range or max value for each slice/block, allowing users to specify a threshold or select the top $x$\% as ROIs.

For example, in the Nyx cosmology dataset, the ROI is the points with a value larger than 81.66 (to serve as the threshold for halo formation~\cite{davis1985evolution,ffis}). Thus, the user can extract ROI using maximum-value thresholding with 81.66. As shown in Fig.~\ref{fig:roi}, by extracting just 0.69\% of the dataset, our solution can capture all the halos, highlighting the importance of random-access decompression.
\begin{table}[h]
  \setlength{\columnsep}{0pt}
  \caption{Our tested datasets}
  \label{tab:dataset}
  \centering\sffamily
  \resizebox{\linewidth}{!}{%
    \newcommand{\MRESO}[2]{\makebox[4.5em][l]{#1}\makebox[7em][l]{#2}}

\begin{tabular}{@{} l llrc @{}}
    \toprule
    \textbf{Dataset} & \textbf{Data type}
                     &
    \textbf{Dimensions}
                     &
    \textbf{\begin{tabular}[c]{@{}c@{}}Per-Timestep\\Data Size\end{tabular}} & \textbf{Domain}
    \\ \midrule
    {Nyx}         & {FP32}    & $512^3$           & 512 MB\quad\null & Cosmology \\
    \cmidrule(l){2-5}
    {WarpX}          & {FP64}   & $256^2 \times 2048$  & 1024 MB\quad\null & Accelerator Physics\\
    \cmidrule(l){2-5}
    {Mag.\_Rec.}       & {FP32}    & $512^3$       & 512 MB \quad\null   & Plasma Physics\\
    \cmidrule(l){2-5}
    {Miranda}         & {FP32}    & $1024^3$           & 4096 MB \quad\null & Turbulence\\
  
    \bottomrule
\end{tabular}
  }
\end{table}

\section{Experimental Evaluation}

\label{sec:eva}

\subsection{Experimental Setup}
\textbf{Test platform.}
The test platform is equipped with 256 GB of memory and two 64-core, 2.25 GHz, AMD EPYC 7742 processors.

\textbf{Test datasets.}
We evaluated our solution on four different scientific simulation datasets, as shown in Table~\ref{tab:dataset}: the Nyx cosmology simulation~\cite{nyx}, the WarpX plasma accelerator simulation (2022 Gordon Bell Prize Winner)~\cite{warpx, warpx-gordon}, the Miranda hydrodynamics simulation~\cite{miranda}, and the Magnetic Reconnection simulation~\cite{magnetic_reconnection}.

\textbf{Comparison baseline.}
We compare our STZ with four SOTA compressors: SZ3, ZFP, MGARD-X (accelerated MGARD), and SPERR, using default settings. Our STZ is the only solution that supports both progressive decompression and random-access decompression. Other compressors support at most one of these features, while SZ3 supports neither. Please refer to Table~\ref{tab:first} for details.

\subsection{Evaluation on Rate-Distortion}
\label{sec:eva-rd}
We plot the rate-distortion curves to compare compression quality, measured by the PSNR, at the same compression ratio (CR) across different compressors.
As shown in Figure~\ref{fig:all-rd} (top-right indicates better performance), our streaming compression solution achieves overall high compression performance in terms of rate-distortion compared to the SOTA solutions.

Specifically, our solution outperforms MGARD-X under all tested datasets and CRs.
Furthermore, our solution yields a significantly better rate-distortion than ZFP. The main reason is that ZFP partitions the data into $4^3$ blocks and processes them separately, which can cause each block to lose its neighbor's spatial information, resulting in low compression quality.

\begin{figure*}[t]
  \centering
  \begin{subfigure}[t]{0.32\linewidth}
    \centering
    \includegraphics[width=\linewidth]{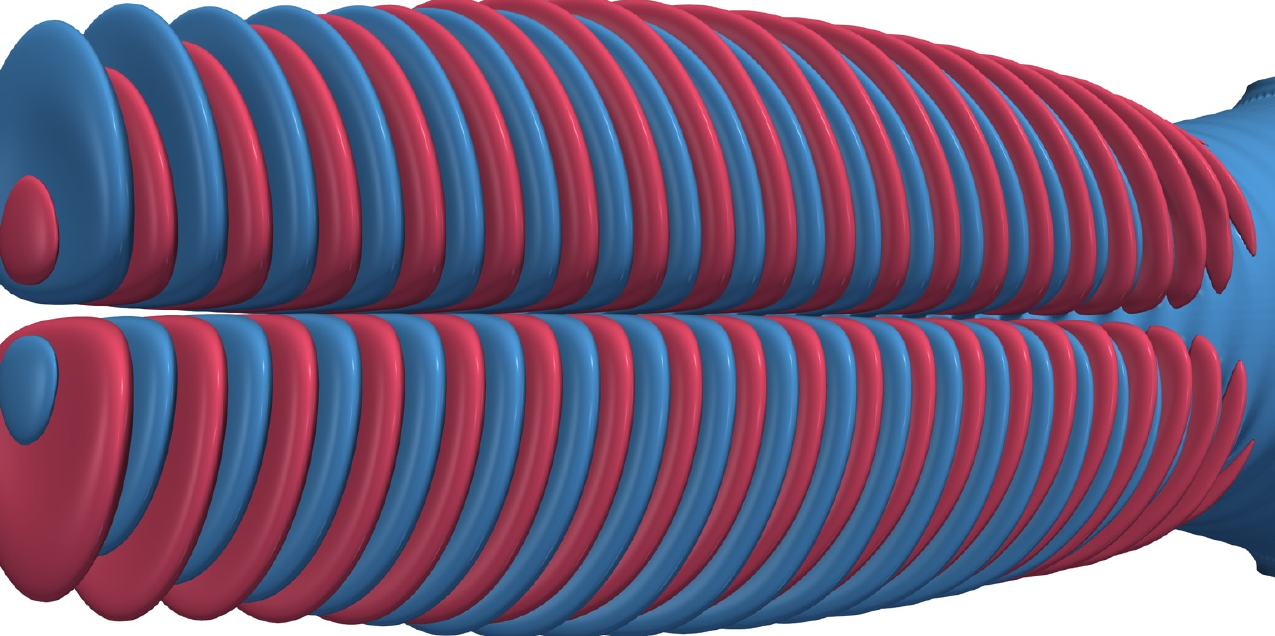}
    \caption[t]{Original data, WarpX}
    \label{fig:vis-wpx-ori}
  \end{subfigure}
  \begin{subfigure}[t]{0.32\linewidth}
    \centering
    \includegraphics[width=\linewidth]{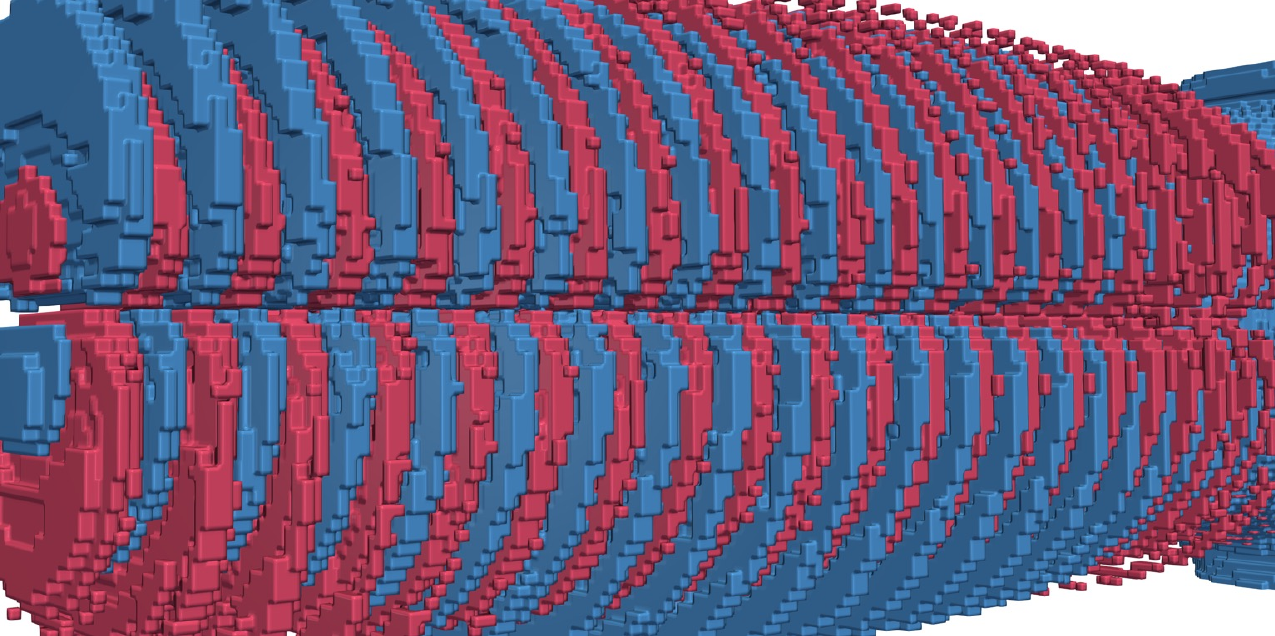}
    \caption{ZFP, SSIM=0.53, PSNR=61, CR=261}
    \label{fig:vis-wpx-zfp}
  \end{subfigure}
  \begin{subfigure}[t]{0.32\linewidth}
    \centering
    \includegraphics[width=\linewidth]{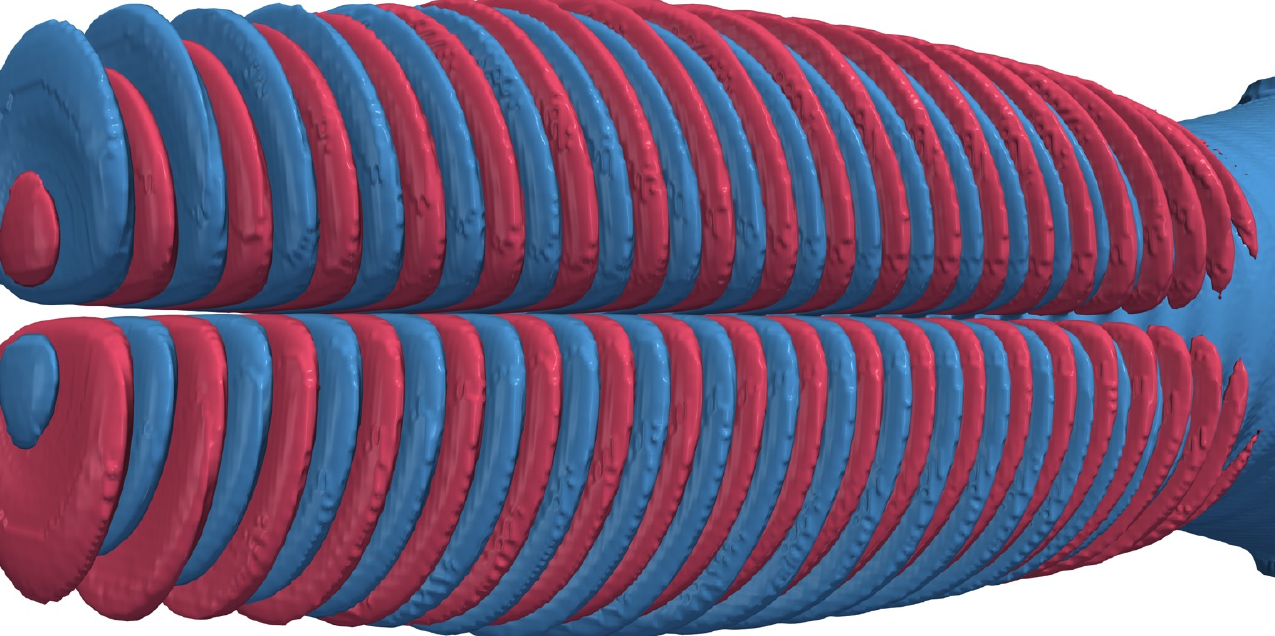}
    \caption{MGARD-X, SSIM=0.85, PSNR=76, CR=296}
    \label{fig:vis-wpx-mgard}
  \end{subfigure}
  \begin{subfigure}[t]{0.32\linewidth}
    \centering
    \includegraphics[width=\linewidth]{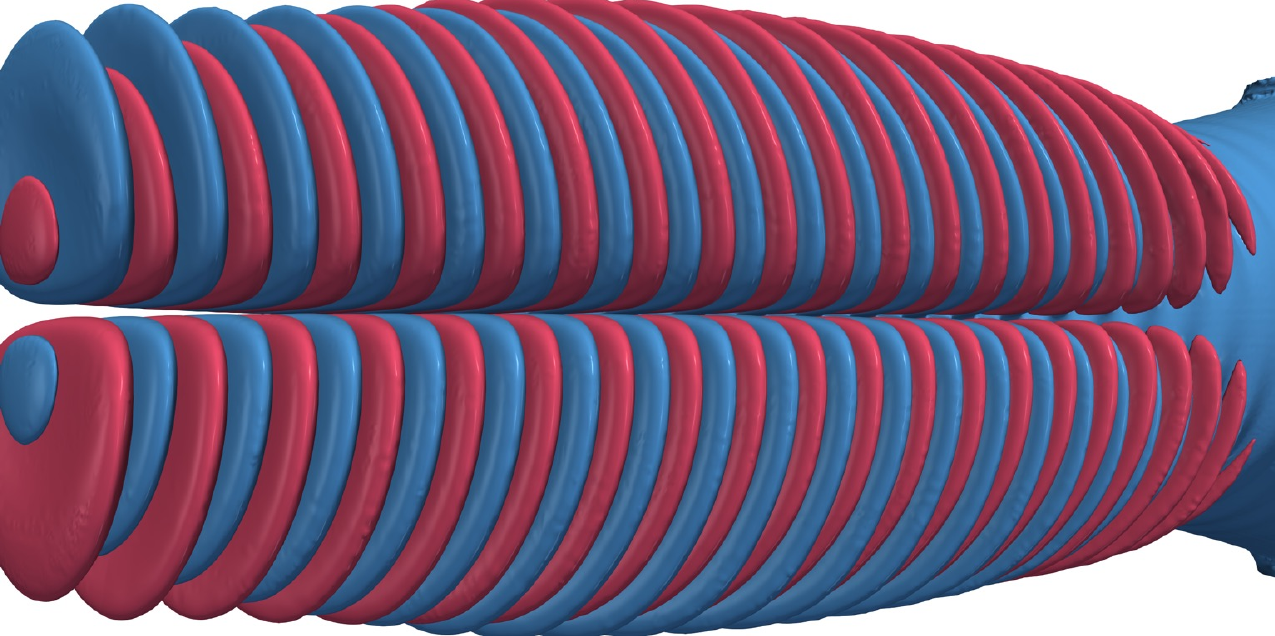}
    \caption{SZ3, SSIM=0.98, PSNR=96.8, CR=297}
    \label{fig:vis-wpx-sz3}
  \end{subfigure}
  \begin{subfigure}[t]{0.32\linewidth}
    \centering
    \includegraphics[width=\linewidth]{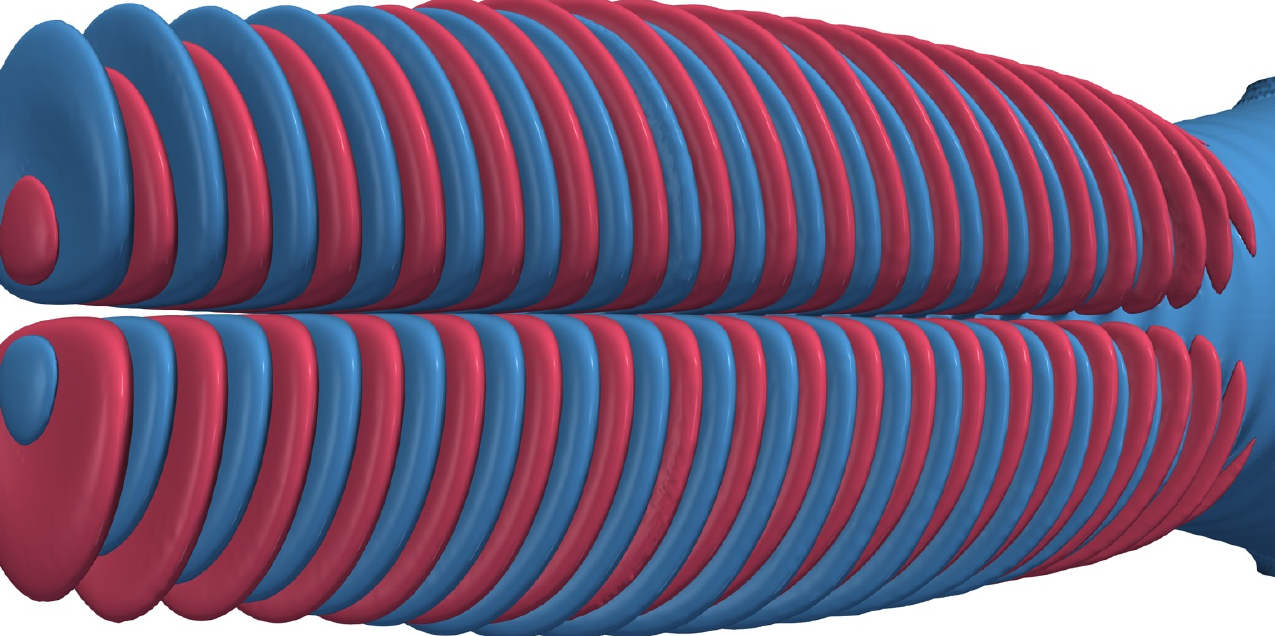}
    \caption{SPERR, SSIM=0.98, PSNR=96.1, CR=295}
    \label{fig:vis-wpx-sperr}
  \end{subfigure}
  \begin{subfigure}[t]{0.32\linewidth}
    \centering
    \includegraphics[width=\linewidth]{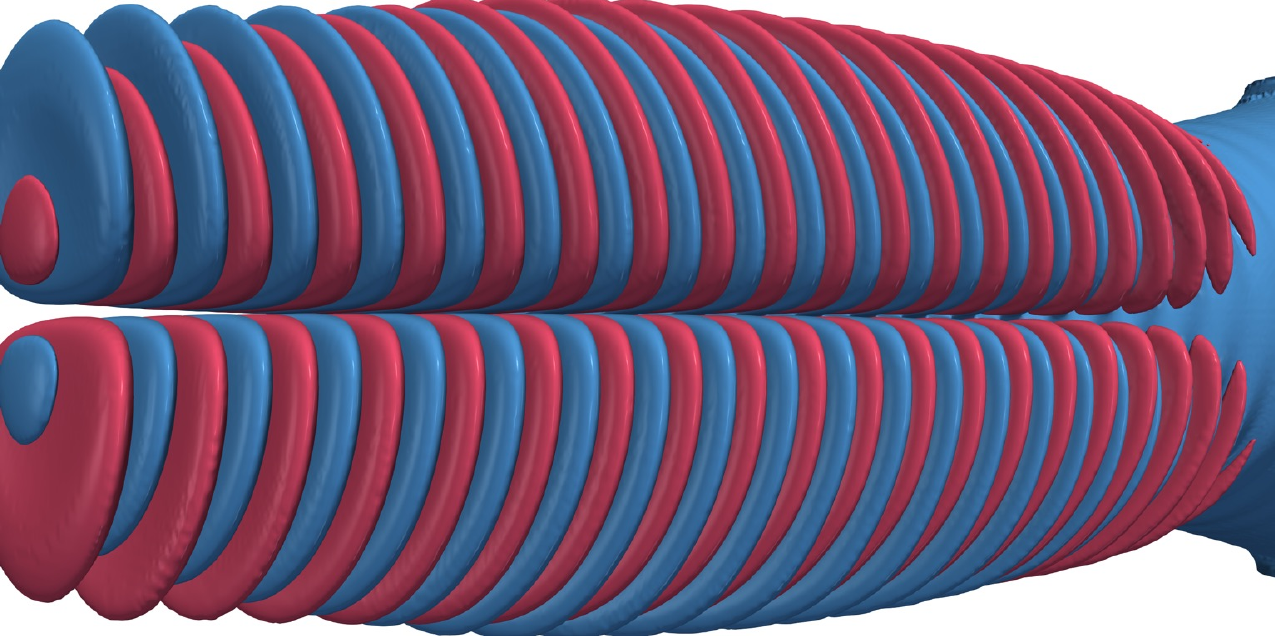}
    \caption{Ours, SSIM=0.99, PSNR=96.5, CR=297}
    \label{fig:vis-wpx-ours}
  \end{subfigure}
  \begin{subfigure}[t]{0.32\linewidth}
    \centering
    \includegraphics[width=\linewidth]{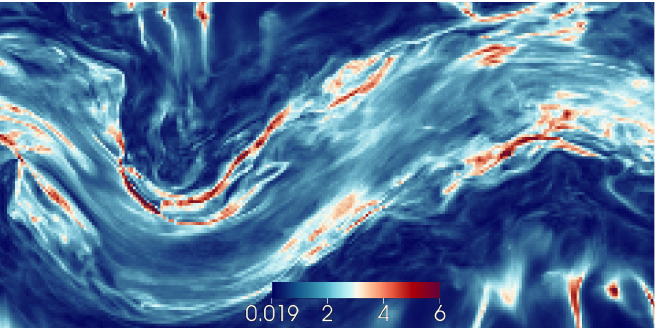}
    \caption[t]{Original data, Magnetic Reconnection}
    \label{fig:vis-mag-ori}
  \end{subfigure}
  \begin{subfigure}[t]{0.32\linewidth}
    \centering
    \includegraphics[width=\linewidth]{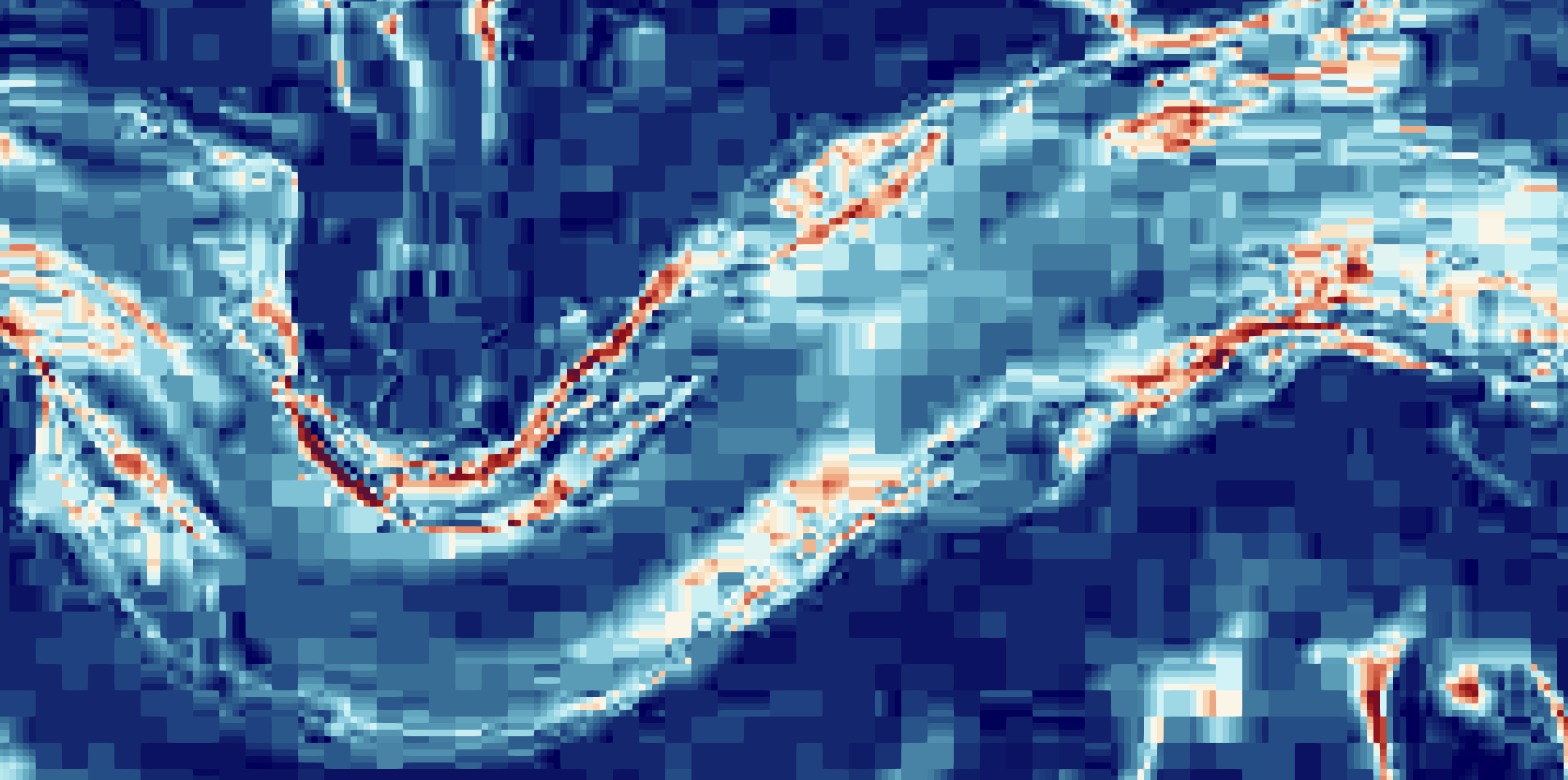}
    \caption{ZFP, SSIM=0.63, PSNR=46, CR=194}
    \label{fig:vis-mag-zfp}
  \end{subfigure}
  \begin{subfigure}[t]{0.32\linewidth}
    \centering
    \includegraphics[width=\linewidth]{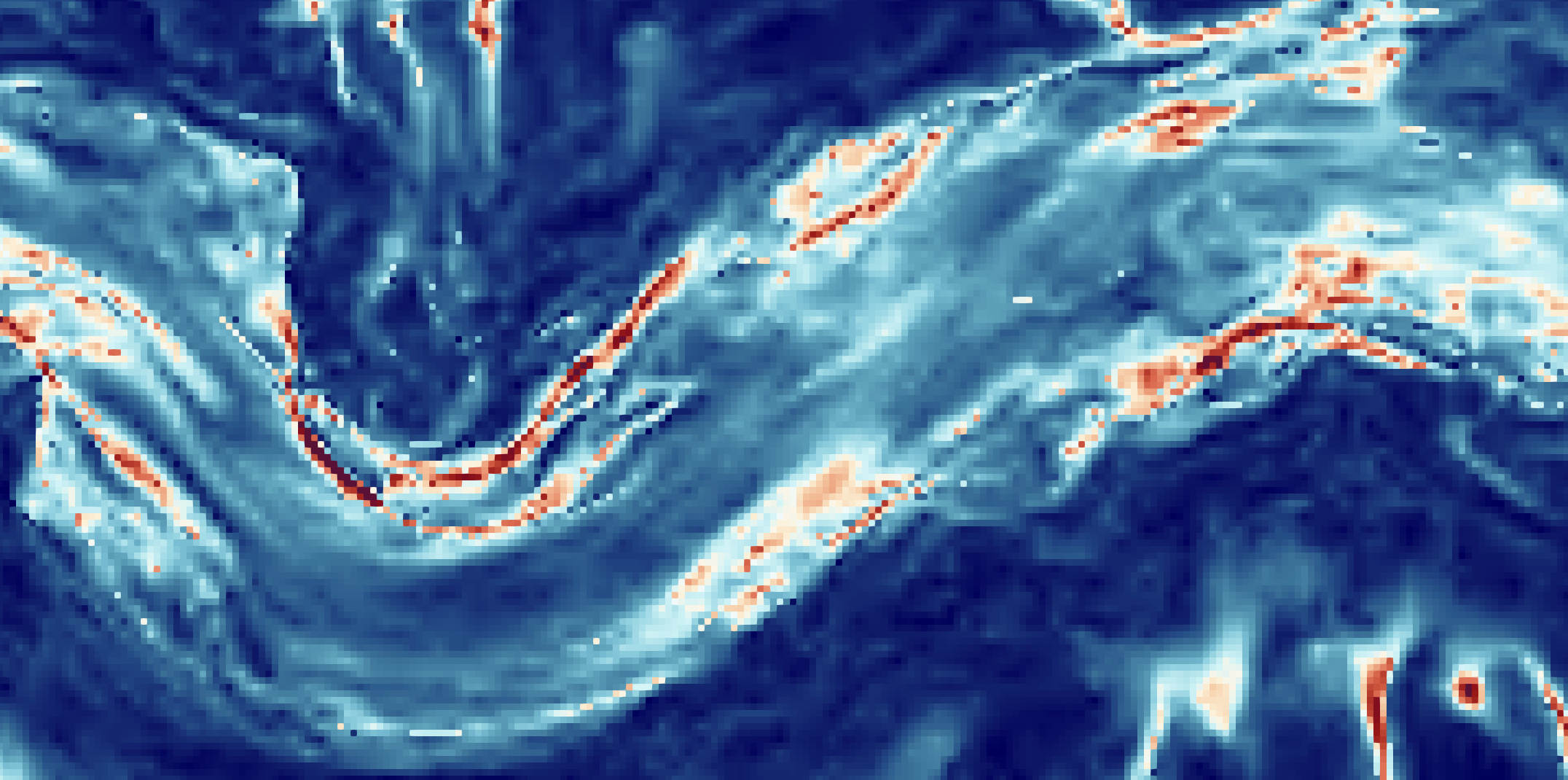}
    \caption{MGARD-X, SSIM=0.79, PSNR=51.2, CR=215}
    \label{fig:vis-mag-mgard}
  \end{subfigure}
  \begin{subfigure}[t]{0.32\linewidth}
    \centering
    \includegraphics[width=\linewidth]{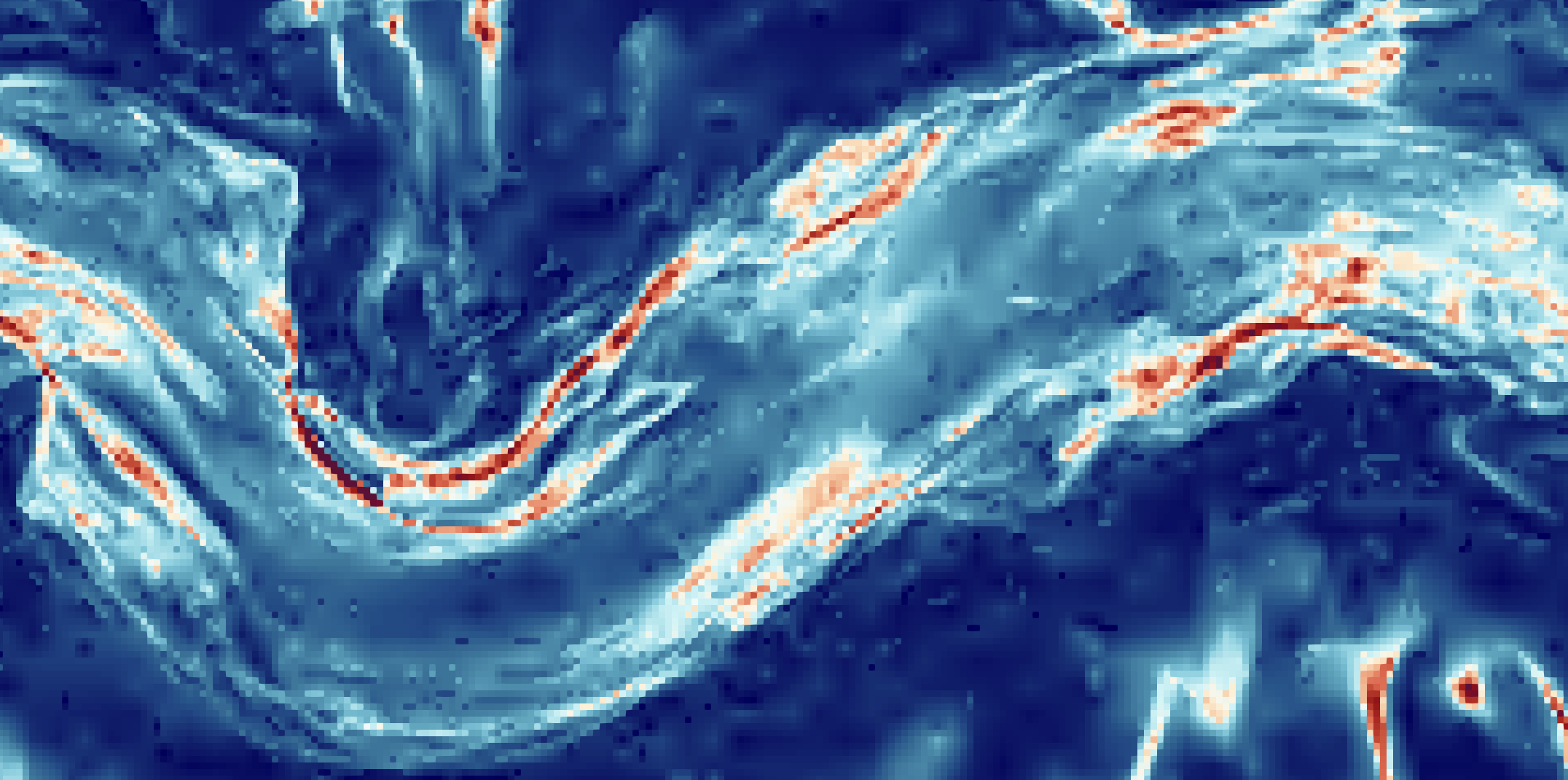}
    \caption{SZ3, SSIM=0.83, PSNR=51.6, CR=215}
    \label{fig:vis-mag-sz3}
  \end{subfigure}
  \begin{subfigure}[t]{0.32\linewidth}
    \centering
    \includegraphics[width=\linewidth]{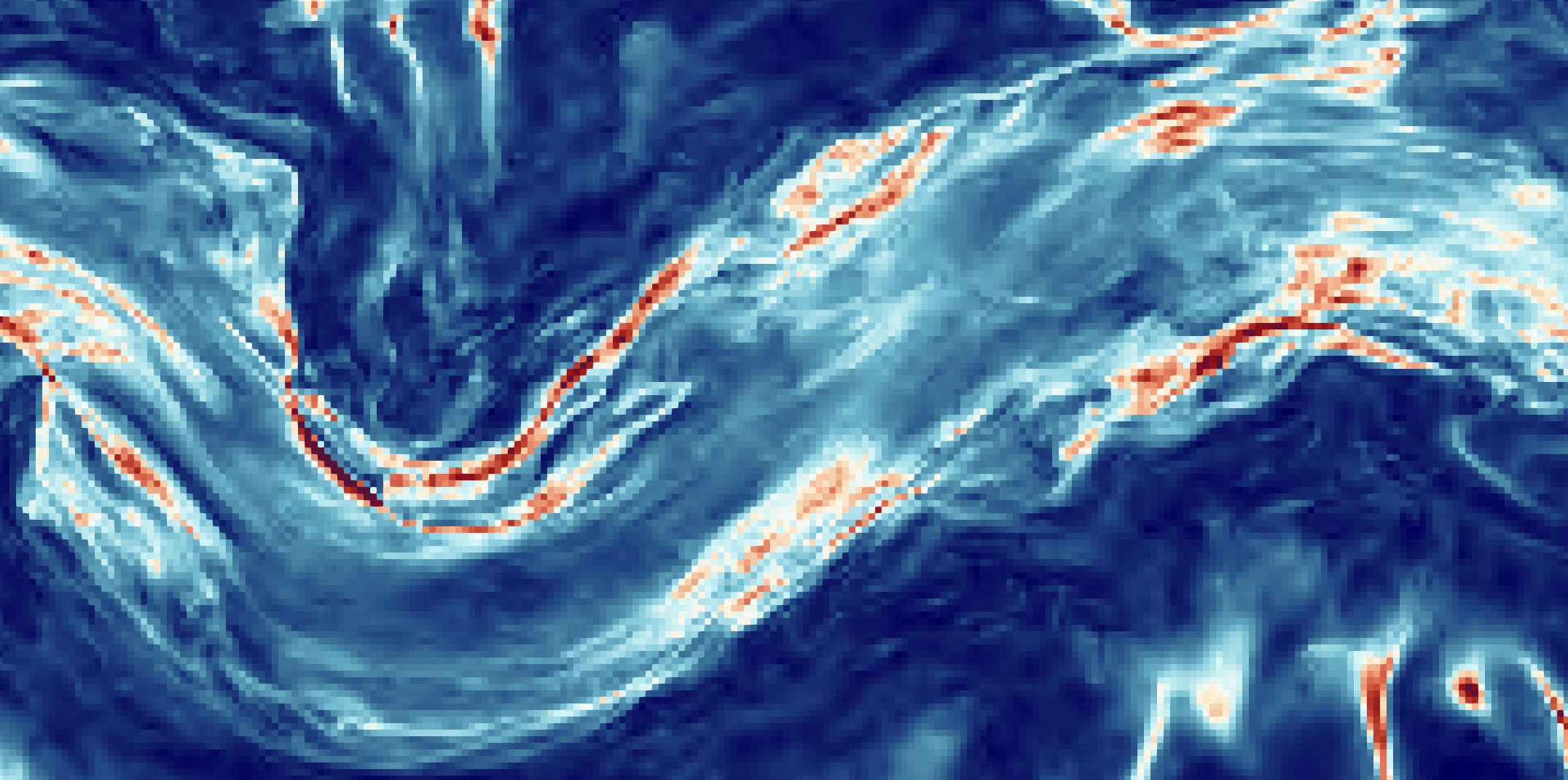}
    \caption{SPERR, SSIM=0.89, PSNR=57.8, CR=215}
    \label{fig:vis-mag-sperr}
  \end{subfigure}
  \begin{subfigure}[t]{0.32\linewidth}
    \centering
    \includegraphics[width=\linewidth]{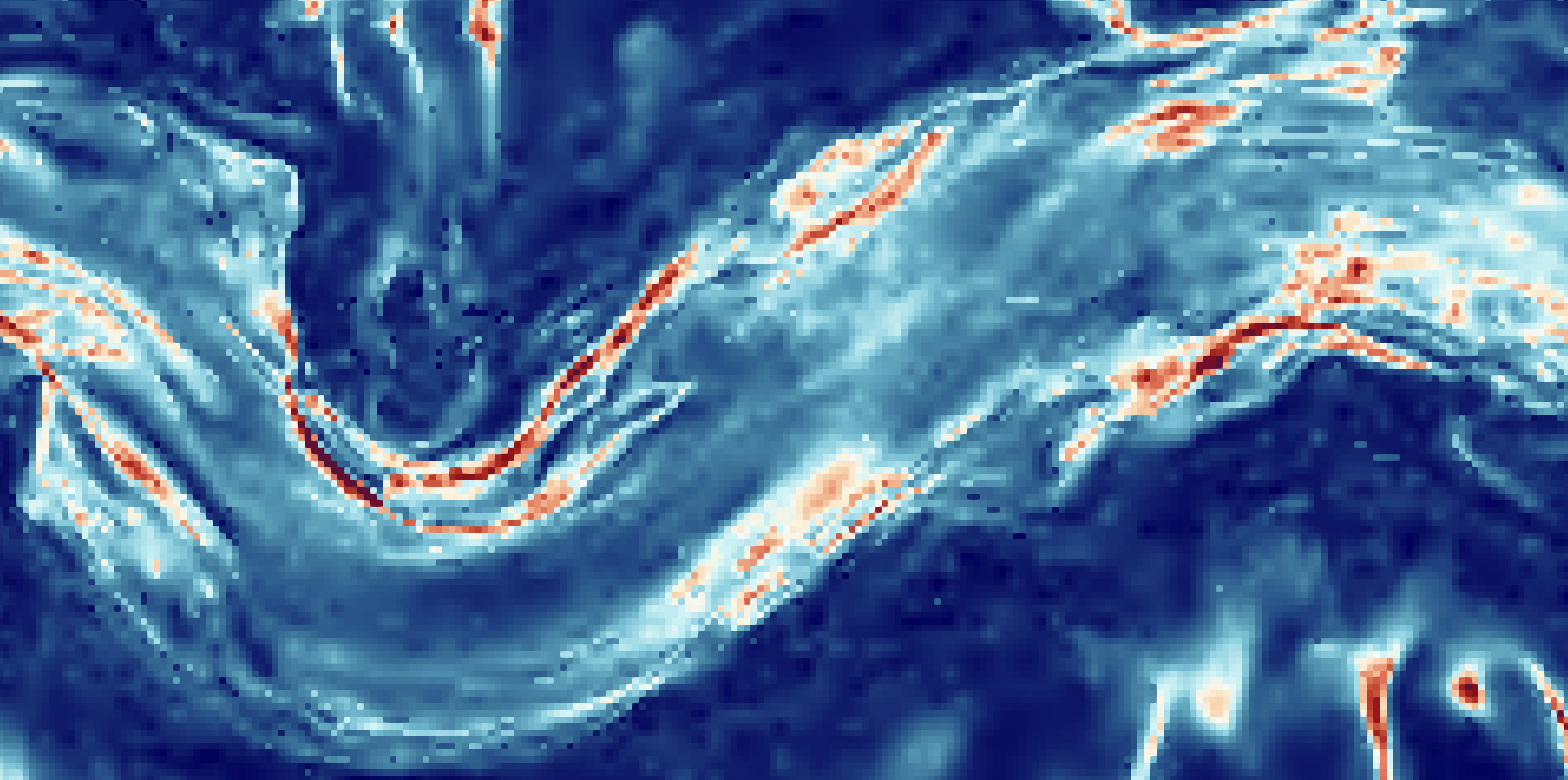}
    \caption{Ours, SSIM=0.83, PSNR=52.4, CR=215}
    \label{fig:vis-mag-ours}
  \end{subfigure}
  \caption[t]{Visual comparison (iso-surface and zoom in 2D slice) of the original data and decompressed data produced by baseline compressors and our STZ at similar compression ratios (ZFP's CRs are lower than the others), on the WarpX and Magnetic Reconnection datasets. For the Magnetic Reconnection dataset, the “Cool to Warm (Extended)” colormap in ParaView is used.}
  \label{fig:vis-mag-wpx}
\end{figure*}

Compared to SZ3, our solution achieves similar rate-distortion performance. Although both our solution and SZ3 are based on interpolation, our method partitions the data to enable streaming, which breaks spatial locality and may reduce compression quality. However, we apply multi-dimensional interpolation, while SZ3 uses only 1D interpolation. Our multidimensional approach captures more spatial redundancy, thereby compensating for the loss caused by partitioning and enabling streaming compression without sacrificing data quality.
Specifically, we achieve slightly better reconstruction quality on Nyx and Magnetic reconnection than SZ3 when the CR is relatively high, as shown in Figure~\ref{fig:all-rd}, and slightly lower data quality on WarpX and Miranda when the CR is low.

Compared to SPERR, our work achieves better quality across all CRs on Nyx and provides similar rate-distortion performance on WarpX at high CR. However, SPERR outperforms our solution on the Magnetic Reconnection and Miranda. Our understanding is that its global wavelet transform more effectively captures widespread high-frequency components for these two datasets (at the cost of slower speed, which will be shown in \S\ref{sec:eva-speed}). The overall compression quality at the same CR is summarized in the last row of Table~\ref{tab:first}.

In Figure~\ref{fig:vis-mag-wpx}, we further compare the visual quality (including SSIM~\cite{ssim}, computed in image space) of the reconstructed data produced by different compressors at similar CRs (ZFP's CRs are lower than the others) on the WarpX and Magnetic Reconnection datasets. The visual quality is consistent with the rate-distortion performance.
Specifically, compared to MGARD (Figures~\ref{fig:vis-wpx-mgard} and~\ref{fig:vis-mag-mgard}), our solution (Figures~\ref{fig:vis-wpx-ours} and~\ref{fig:vis-mag-ours}) provides better visual quality on WarpX and slightly better visual quality on the Magnetic Reconnection dataset.
Compared to SZ3 (Figures~\ref{fig:vis-wpx-sz3} and~\ref{fig:vis-mag-sz3}), our solution achieves similar visual quality.
Compared to SPERR (Figures~\ref{fig:vis-wpx-sperr} and~\ref{fig:vis-mag-sperr}), our solution achieves similar visual quality on WarpX and slightly lower visual quality on the Magnetic Reconnection dataset.
Compared to ZFP (Figures~\ref{fig:vis-wpx-zfp} and~\ref{fig:vis-mag-zfp}), our solution provides significantly better visual quality, even at higher CRs. This is because ZFP produces noticeable block-wise compression artifacts due to its block-wise nature, which disrupts spatial coherence.
\begin{table}[h]
  \setlength{\columnsep}{0pt}
  \caption{Compression and decompression times (in seconds) for different compressors under Serial and OMP modes. OMP results are shown in blue. Asterisks (*) indicate that the SZ3 OMP configuration leads to a drop in compression ratio (CR). ZFP and MGARD-X do not support OpenMP acceleration for decompression.
  }
  \label{tab:speed}
  \centering\sffamily
  \resizebox{0.83\linewidth}{!}{%

\begin{tabular}{@{}cclrrrrr@{}}
\toprule
 & & & Ours & SZ3 & SPERR & ZFP & MGARDX \\ \midrule
\multirow{8}{*}{\centering\rotatebox[origin=c]{90}{\makebox[2.5cm][c]{Compression}}}
  & \multirow{2}{*}{NYX} & Serial & 2.1 & 5.0 & 44.6 & 1.3 & 15.7 \\
  & & OMP & \textcolor{blue!40!cyan}{0.3} & \textcolor{blue!40!cyan}{1.3*} & \textcolor{blue!40!cyan}{6.5} & \textcolor{blue!40!cyan}{0.3} & \textcolor{blue!40!cyan}{5.3} \\ \cmidrule(l){2-8}
  & \multirow{2}{*}{Mag.\_Rec.} & Serial & 2.1 & 5.0 & 32.8 & 1.1 & 17.4 \\
  & & OMP & \textcolor{blue!40!cyan}{0.3} & \textcolor{blue!40!cyan}{1.4} & \textcolor{blue!40!cyan}{5.8} & \textcolor{blue!40!cyan}{0.3} & \textcolor{blue!40!cyan}{4.8} \\ \cmidrule(l){2-8}
  & \multirow{2}{*}{WarpX} & Serial & 2.5 & 4.0 & 44.3 & 2.1 & 17.9 \\
  & & OMP & \textcolor{blue!40!cyan}{0.5} & \textcolor{blue!40!cyan}{0.6} & \textcolor{blue!40!cyan}{6.4} & \textcolor{blue!40!cyan}{0.4} & \textcolor{blue!40!cyan}{5.8} \\ \cmidrule(l){2-8}
  & \multirow{2}{*}{Miranda} & Serial & 18.5 & 44.9 & 350.4 & 16.3 & 175.4 \\
  & & OMP & \textcolor{blue!40!cyan}{3.5} & \textcolor{blue!40!cyan}{11.4*} & \textcolor{blue!40!cyan}{49.4} & \textcolor{blue!40!cyan}{2.8} & \textcolor{blue!40!cyan}{42.4} \\ \midrule
\multirow{8}{*}{\centering\rotatebox[origin=c]{90}{\makebox[2.5cm][c]{Decompression}}}
  & \multirow{2}{*}{NYX} & Serial & 1.1 & 7.2 & 40.6 & 0.6 & 15.4 \\
  & & OMP & \textcolor{blue!40!cyan}{0.3} & \textcolor{blue!40!cyan}{2.0*} & \textcolor{blue!40!cyan}{5.8} & N/A & N/A \\ \cmidrule(l){2-8}
  & \multirow{2}{*}{Mag.\_Rec.} & Serial & 1.2 & 7.2 & 34.6 & 0.5 & 14.9 \\
  & & OMP & \textcolor{blue!40!cyan}{0.3} & \textcolor{blue!40!cyan}{2.0} & \textcolor{blue!40!cyan}{5.7} & N/A & N/A \\ \cmidrule(l){2-8}
  & \multirow{2}{*}{WarpX} & Serial & 1.5 & 5.6 & 40.2 & 1.0 & 18.3 \\
  & & OMP & \textcolor{blue!40!cyan}{0.4} & \textcolor{blue!40!cyan}{0.8} & \textcolor{blue!40!cyan}{5.1} & N/A & N/A \\ \cmidrule(l){2-8}
  & \multirow{2}{*}{Miranda} & Serial & 11.7 & 63.1 & 327.3 & 6.7 & 132.6 \\
  & & OMP & \textcolor{blue!40!cyan}{2.9} & \textcolor{blue!40!cyan}{17.0*} & \textcolor{blue!40!cyan}{52.7} & N/A & N/A \\ \bottomrule
\end{tabular}

  }
\end{table}

\subsection{Evaluation on Speed}
\label{sec:eva-speed}

The compression and decompression times using different compressors on different datasets are shown in Table~\ref{tab:speed}.
To ensure fair speed comparisons, we selected error bounds that produce similar compression quality for SZ3 and our solution. In addition, SZ3, ZFP, SPERR, and MGARD-X were evaluated using the same error bound. Both serial and OpenMP modes were evaluated, utilizing eight OpenMP threads. ZFP and MGARD-X do not support OpenMP acceleration for decompression.
Additionally, SZ3's OpenMP mode can result in a reduced compression ratio compared to its serial mode, with up to 4\% lower compression ratio on the Nyx dataset and up to 40\% lower on the Miranda dataset.

As shown in Table~\ref{tab:speed}, in serial mode, our solution achieves the second-highest compression and decompression speeds among all baselines, only slower than ZFP, which is specifically designed for high-speed compression but offers low compression quality, as discussed in \S\ref{sec:eva-rd}. Compared to SZ3, SPERR, and MGARD-X, our solution achieves compression speedups of up to 2.5$\times$, 21$\times$, and 9.5$\times$, respectively. Additionally, our solution achieves decompression speedups of up to 6.5$\times$, 37$\times$, and 14$\times$ compared to SZ3, SPERR, and MGARD-X, respectively.

In OpenMP mode, our solution achieves even higher compression speedups, as our low data dependency design provides high parallel efficiency.
We achieve up to 4.7$\times$, 22$\times$, and 18$\times$ compression speedups over SZ3, SPERR, and MGARD-X. Our compression speed is also comparable to ZFP when using OpenMP.
For decompression, we achieve speedups of up to 6.7$\times$ and 19$\times$ over SZ3 and SPERR.
Moreover, since our solution supports OpenMP acceleration for decompression while ZFP does not, we achieve overall higher compression and decompression speeds than ZFP in OpenMP mode.

\subsection[Discussion on SZ3 vs {\thiswork} Methods]{Discussion on the Different Interpolation Methods Used by SZ3 and \thiswork}
Although both SZ3 and our STZ are based on interpolation prediction, there are three key differences in the interpolation.

\textbf{Multi-dimensional interpolation.} First, SZ3 employs only 1D interpolation, while our solution uses multi-dimensional interpolation. Our approach leverages more spatial information and mitigates the negative effects introduced by hierarchical partitioning, which is necessary to support streaming compression. Thus, our method achieves a compression quality similar to SZ3 while supporting random-access decompression and progressive decompression.

\textbf{Better caching efficiency.} Second, our prediction method has better data locality. Our prediction method uses nearby data points to perform multi-dimensional interpolation, which is more cache-friendly. In contrast, SZ3 performs long-range 1D interpolation. For example, for data with a size of $N$, SZ3 will use points $i_{0}$ and $i_{N}$ to interpolate $i_{\frac{N}{2}}$, resulting in large memory jumps across the data array and more cache misses.
In contrast, our approach accesses data locally and can process the entire data array sequentially in a cache-aligned manner, leading to significantly better cache performance, thus higher compression and decompression speed.

\textbf{Less data dependency.} Last, our prediction involves less data dependency. Prediction-based compressors like SZ3 must use decompressed data for predictions to avoid error propagation and ensure that errors remain within the error bound. In SZ3, the interpolation process creates strong dependencies among data points. At least half of the data points are used to predict others, so these points must be decompressed during compression to avoid error propagation, which incurs high overhead.
In contrast, in our solution, all data points in level 3 depend solely on levels 1 and 2, with no dependencies among themselves. Level 3 accounts for $87.5\%$ of the 3D data, and since no data point is dependent on level 3, $87.5\%$ of the data does not require decompression during compression. This significantly reduces decompression overhead compared to SZ3 and improves compression speed. Moreover, reduced data dependency improves parallel efficiency, allowing our solution to achieve even higher speedup than SZ3 in OpenMP mode, as mentioned in \S\ref{sec:eva-speed}.

\begin{figure*}[t]
  \centering
  \begin{subfigure}[t]{0.24\linewidth}
    \centering
    \includegraphics[width=\linewidth]{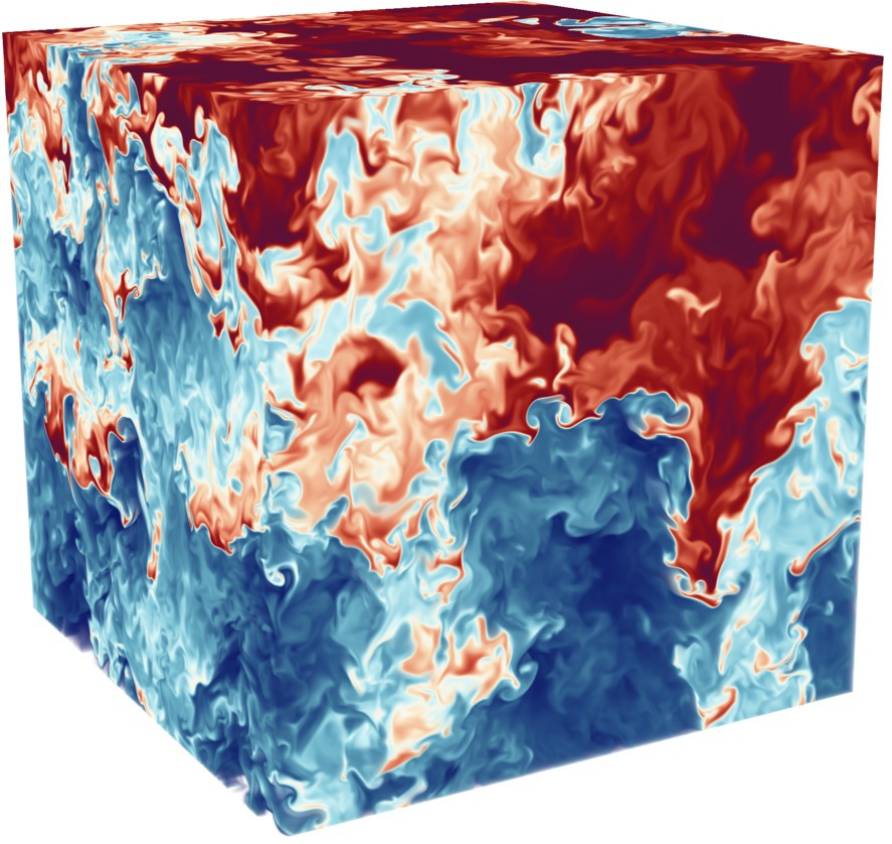}
    \caption[t]{Original data, $1024^3$}
    \label{fig:vis-ori}
  \end{subfigure}~%
  \begin{subfigure}[t]{0.24\linewidth}
    \centering
    \includegraphics[width=\linewidth]{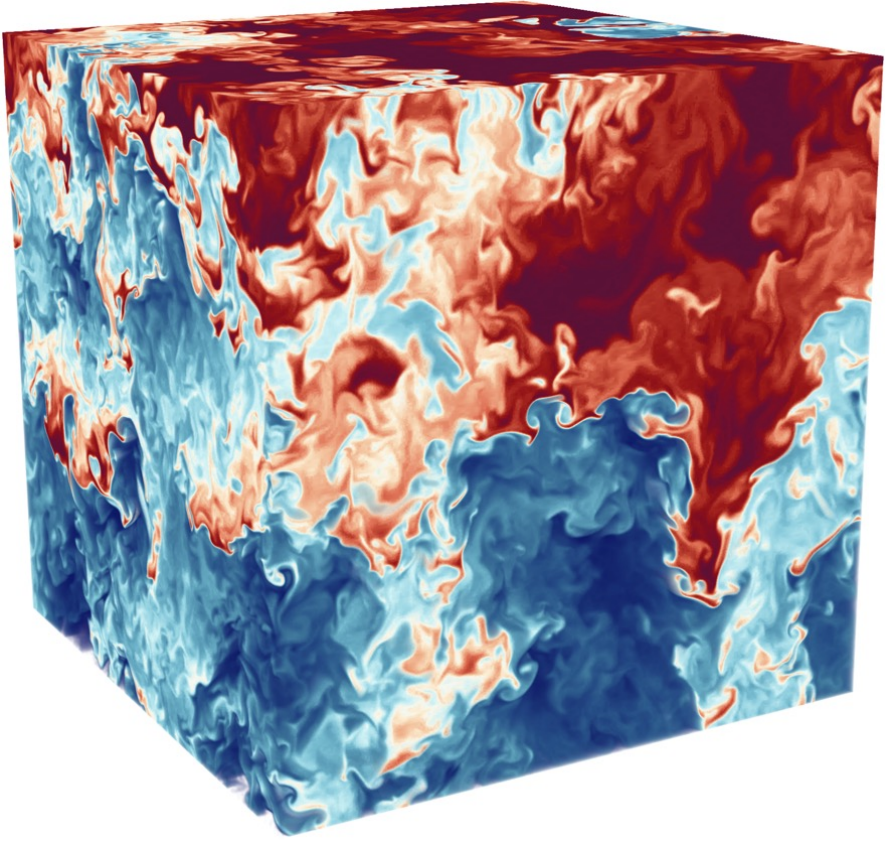}
    \caption{$1024^3$, SSIM=0.96, De. time=11.4s}
    \label{fig:vis-full}
  \end{subfigure}~%
  \begin{subfigure}[t]{0.24\linewidth}
    \centering
    \includegraphics[width=\linewidth]{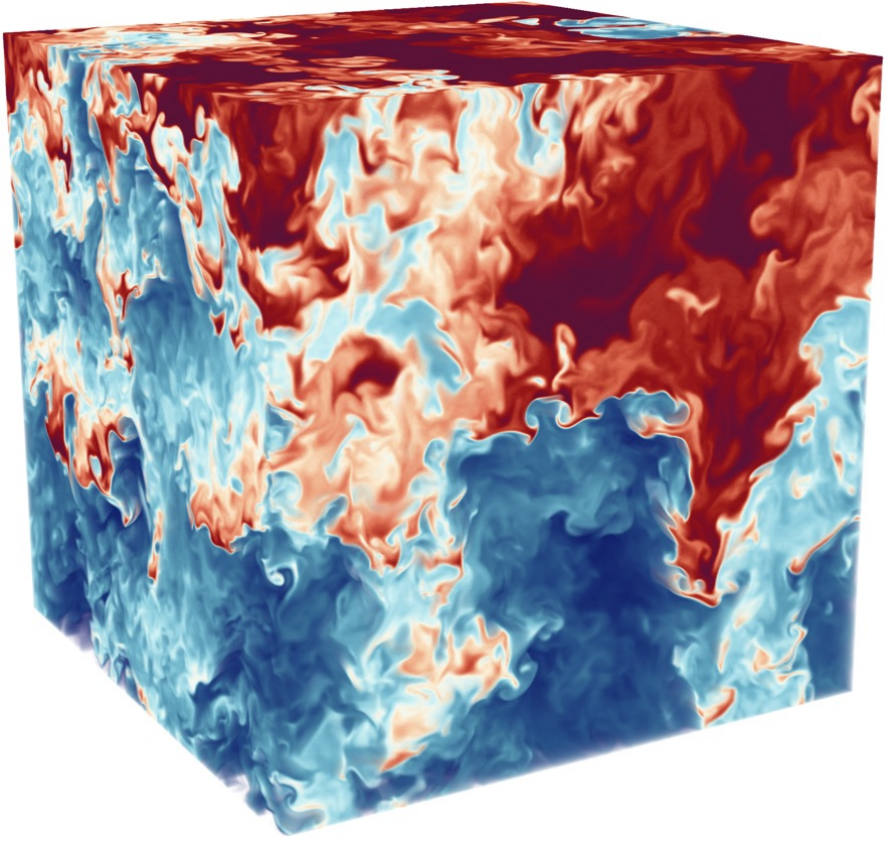}
    \caption{$512^3$, SSIM=0.86, De. time=2.5s}
    \label{fig:vis-mid}
  \end{subfigure}~%
  \begin{subfigure}[t]{0.24\linewidth}
    \centering
    \includegraphics[width=\linewidth]{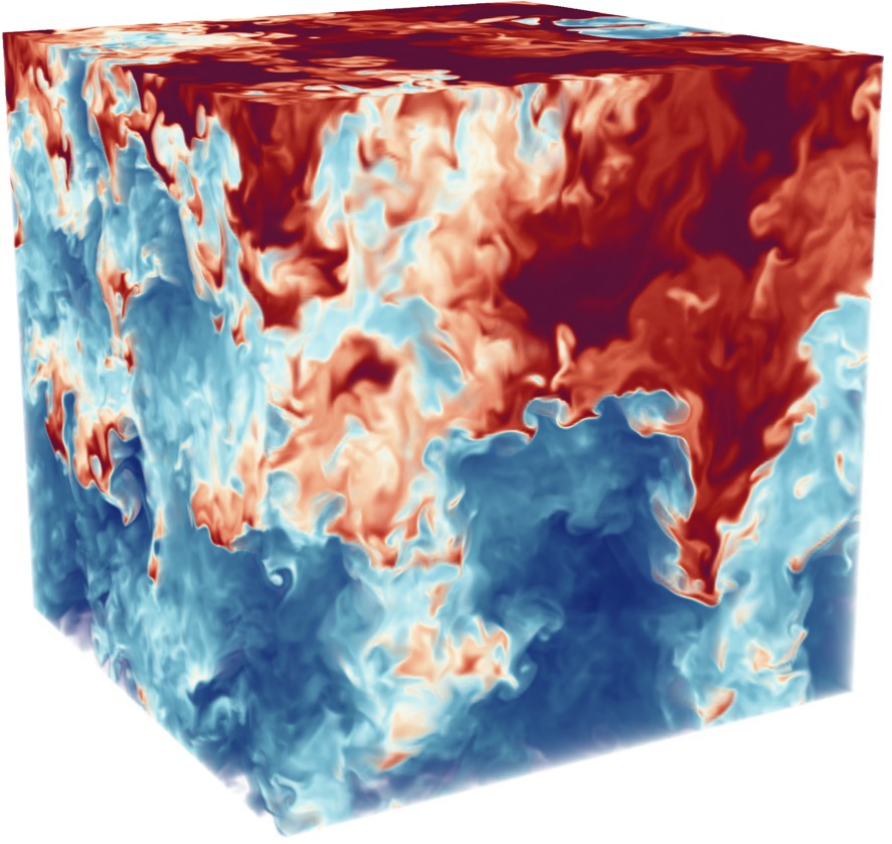}
    \caption{$256^3$, SSIM=0.74, De. time=0.71s}
    \label{fig:vis-low}
  \end{subfigure}
  \caption[t]{Volume rendering of the original data of the Miranda dataset and progressively decompressed data produced by our STZ at different resolutions. The CR of the full-resolution data is 447, and the corresponding decompression times (in seconds) are shown for each resolution.}
  \label{fig:vis-prog}
\end{figure*}

\subsection{Evaluation on Streaming Decompression}
\label{sec:eva-str}
We now demonstrate the streaming decompression capability of our solution, which includes both progressive decompression and random-access decompression.

\textbf{Progressive decompression.}
Figure~\ref{fig:vis-prog} shows the visualization of the original data of the Miranda dataset, as well as the progressive reconstruction data at full ($1024^3$), coarse ($512^3$), and the coarsest ($256^3$) resolution. At the coarsest and coarse levels, some of the details were lost due to down-sampling, however, the overall structure remains clear even for the coarsest resolution. This feature allows users to grasp the overall structure and efficiently identify ROI for selective high-resolution reconstruction.

Moreover, the decompression time for the coarsest and coarse levels is much shorter than that of the full-resolution. Progressive decompression can also greatly reduce memory usage during decompression and visualization, and shorten the visualization time.

\textbf{Random access decompression.}
Table~\ref{tab:rd} presents the decompression time for the full dataset, for random-access decompression of a $100^3$ ROI box, and to a $1024^2$ 2D slice of the Miranda dataset. As shown in the last column, our solution achieves overall decompression time savings of up to 67.5\% for 3D box access and up to 82.5\% for 2D slice access.
\begin{table}[h]
  \setlength{\columnsep}{0pt}
  \caption{Time breakdown (in sec) to decompress all data, random-access one 3D ROI box, and one 2D slice in the Miranda dataset. The blue text indicates the stages that can achieve savings in random-access decompression.}
  \label{tab:rd}
  \centering\sffamily
  \resizebox{\linewidth}{!}{%
    \begin{tabular}{@{}lcccccccr@{}}
\toprule
Case  & L1 SZ3  & L2 dec. & L2 pre. & L2 rec. & L3 dec. & L3 pre. & L3 rec. & Sum \\ \midrule
All   & 0.71 & 0.44    & 0.64    & 0.68    & 2.64    & 4.83    & 1.80    & 11.7 \\
Box   & 0.71 & 0.44    & \textcolor{blue!40!cyan}{3E-3}    & \textcolor{blue!40!cyan}{3E-4}    & 2.64    & \textcolor{blue!40!cyan}{2E-2}    & \textcolor{blue!40!cyan}{4E-3}    & \textcolor{blue!40!cyan}{3.8} \\
Slice & 0.71 & \textcolor{blue!40!cyan}{0.18}    & \textcolor{blue!40!cyan}{1E-3}    & \textcolor{blue!40!cyan}{2E-4}    & \textcolor{blue!40!cyan}{1.15}    & \textcolor{blue!40!cyan}{7E-3}    & \textcolor{blue!40!cyan}{2E-3}    & \textcolor{blue!40!cyan}{2.1} \\ \bottomrule
\end{tabular}
  }
\end{table}

Our decompression process includes multiple steps. First, SZ3 decompresses the coarsest level 1 (first column, L1 SZ3). Then, the prediction errors of level 2 are decoded (L2 dec.), and level 1 is used to predict and restore level 2 with these errors (L2 pre.). Levels 1 and~2 are then reassembled to reconstruct a coarse representation of the data (L2 rec.), which is the reverse of the partitioning performed during compression (Figure~\ref{fig:split}). Finally, level 3 follows the same procedure, using levels 1 and~2 for prediction and reconstructs the full-resolution data.

As shown in Table~\ref{tab:rd}, when decompressing either a 3D ROI box or a 2D slice, the prediction and reconstruction stages for levels 2 and 3 (L2/3 pre. and L2/3 rec.) achieve nearly 100\%  savings compared to prediction and reconstruction of the entire dataset. This efficiency comes from our method’s ability to avoid intra-level dependencies in prediction, allowing computing only the ROI during prediction and reconstruction while not touching the remaining data.

As for the decoding stages at levels 2 and 3 (i.e., L2/3 dec.),
there exist dependencies because the data points within each sub-block are encoded together. This dependency necessitates decoding the entire sub-block to access specific data points. As discussed in \S\ref{sec:rd-acc}, random-access decompression of a 3D ROI box requires decoding all relevant sub-blocks, resulting in no decoding time savings. In contrast, for random-access decompression of a 2D slice, only a subset of sub-blocks (e.g., 3 out of 7 in level 3) needs to be decoded, leading to up to 57\% savings in decoding time.

\section{Conclusion and Future Work}
We propose a streaming compression framework that enables on-demand data access and flexible analysis workflows. While all SOTA scientific lossy compressors support at most one of progressive decompression or random-access decompression, they come with significant sacrifices in compression quality or speed. Our work supports both streaming decompression features and provides high compression quality and high speed. Evaluation against four SOTA lossy compressors on four scientific datasets shows that our work achieves the best overall trade-off between compression quality and speed. Specifically, our work achieves similar compression quality and up to $4.7\times$ compression speed and $6.7\times$ decompression speed compared to the non-streaming compressor SZ3.

In the future, we will improve the speed of our work through code optimization and leverage GPUs for further acceleration.
We also plan to dynamically incorporate other cubic spline polynomial constraints (currently using only the not-a-knot constraint) to improve compression quality.
In addition, we intend to enable random-access Huffman decoding to further reduce the overhead in random-access decompression."

\begin{acks}

  This work was supported by the U.S. Department of Energy through the Los Alamos National Laboratory. Los Alamos National Laboratory is operated by Triad National Security, LLC, for the National Nuclear Security Administration of U.S. Department of Energy (Contract No. 89233218CNA000001).
  This work was also supported in part by the National Science Foundation under Grant Numbers 2311876, 2326495, 2247060, 2247080, 2344717, and 2514035.
  This work was also supported by the U.S. Department of Energy (DOE) RAPIDS-2 SciDAC project under contract number DE-AC05-00OR22725.

\end{acks}

\bibliographystyle{ACM-Reference-Format}
\bibliography{refs}


\begin{thebibliography}{49}

%%% the Web address.

\ifx \showCODEN    \undefined \def \showCODEN     #1{\unskip}     \fi
\ifx \showDOI      \undefined \def \showDOI       #1{#1}\fi
\ifx \showISBNx    \undefined \def \showISBNx     #1{\unskip}     \fi
\ifx \showISBNxiii \undefined \def \showISBNxiii  #1{\unskip}     \fi
\ifx \showISSN     \undefined \def \showISSN      #1{\unskip}     \fi
\ifx \showLCCN     \undefined \def \showLCCN      #1{\unskip}     \fi
\ifx \shownote     \undefined \def \shownote      #1{#1}          \fi
\ifx \showarticletitle \undefined \def \showarticletitle #1{#1}   \fi
\ifx \showURL      \undefined \def \showURL       {\relax}        \fi
% invisible to TeX
\providecommand\bibfield[2]{#2}
\providecommand\bibinfo[2]{#2}
\providecommand\natexlab[1]{#1}
\providecommand\showeprint[2][]{arXiv:#2}

\bibitem[\protect\citeauthoryear{Ainsworth, Tugluk, Whitney, and
  Klasky}{Ainsworth et~al\mbox{.}}{2018}]%
        {ainsworth2018multilevel}
\bibfield{author}{\bibinfo{person}{Mark Ainsworth}, \bibinfo{person}{Ozan
  Tugluk}, \bibinfo{person}{Ben Whitney}, {and} \bibinfo{person}{Scott
  Klasky}.} \bibinfo{year}{2018}\natexlab{}.
\newblock \showarticletitle{Multilevel techniques for compression and reduction
  of scientific data—the univariate case}.
\newblock \bibinfo{journal}{\emph{Computing and Visualization in Science}}
  \bibinfo{volume}{19}, \bibinfo{number}{5--6} (\bibinfo{year}{2018}),
  \bibinfo{pages}{65--76}.
\newblock


\bibitem[\protect\citeauthoryear{Baker, Hammerling, and Turton}{Baker
  et~al\mbox{.}}{2019}]%
        {baker2019evaluating}
\bibfield{author}{\bibinfo{person}{Allison~H Baker}, \bibinfo{person}{Dorit~M
  Hammerling}, {and} \bibinfo{person}{Terece~L Turton}.}
  \bibinfo{year}{2019}\natexlab{}.
\newblock \showarticletitle{Evaluating image quality measures to assess the
  impact of lossy data compression applied to climate simulation data}. In
  \bibinfo{booktitle}{\emph{Computer Graphics Forum}},
  Vol.~\bibinfo{volume}{38}. Wiley Online Library, \bibinfo{pages}{517--528}.
\newblock


\bibitem[\protect\citeauthoryear{Ballester-Ripoll, Lindstrom, and
  Pajarola}{Ballester-Ripoll et~al\mbox{.}}{2020}]%
        {ballester2019tthresh}
\bibfield{author}{\bibinfo{person}{Rafael Ballester-Ripoll},
  \bibinfo{person}{Peter Lindstrom}, {and} \bibinfo{person}{Renato Pajarola}.}
  \bibinfo{year}{2020}\natexlab{}.
\newblock \showarticletitle{TTHRESH: Tensor Compression for Multidimensional
  Visual Data}.
\newblock \bibinfo{journal}{\emph{IEEE Transactions on Visualization and
  Computer Graphics}} \bibinfo{volume}{26}, \bibinfo{number}{9}
  (\bibinfo{year}{2020}), \bibinfo{pages}{2891--2903}.
\newblock
\urldef\tempurl%
\url{https://doi.org/10.1109/TVCG.2019.2904063}
\showDOI{\tempurl}


\bibitem[\protect\citeauthoryear{Bhatia, Hoang, Morrical, Pascucci, Bremer, and
  Lindstrom}{Bhatia et~al\mbox{.}}{2022}]%
        {bhatia2022amm}
\bibfield{author}{\bibinfo{person}{Harsh Bhatia}, \bibinfo{person}{Duong
  Hoang}, \bibinfo{person}{Nate Morrical}, \bibinfo{person}{Valerio Pascucci},
  \bibinfo{person}{Peer-Timo Bremer}, {and} \bibinfo{person}{Peter Lindstrom}.}
  \bibinfo{year}{2022}\natexlab{}.
\newblock \showarticletitle{AMM: Adaptive multilinear meshes}.
\newblock \bibinfo{journal}{\emph{IEEE Transactions on Visualization and
  Computer Graphics}} \bibinfo{volume}{28}, \bibinfo{number}{6}
  (\bibinfo{year}{2022}), \bibinfo{pages}{2350--2363}.
\newblock


\bibitem[\protect\citeauthoryear{Cappello, Di, Li, Liang, Gok, Tao, Yoon, Wu,
  Alexeev, and Chong}{Cappello et~al\mbox{.}}{2019}]%
        {cappello2019use}
\bibfield{author}{\bibinfo{person}{Franck Cappello}, \bibinfo{person}{Sheng
  Di}, \bibinfo{person}{Sihuan Li}, \bibinfo{person}{Xin Liang},
  \bibinfo{person}{Ali~Murat Gok}, \bibinfo{person}{Dingwen Tao},
  \bibinfo{person}{Chun~Hong Yoon}, \bibinfo{person}{Xin-Chuan Wu},
  \bibinfo{person}{Yuri Alexeev}, {and} \bibinfo{person}{Frederic~T Chong}.}
  \bibinfo{year}{2019}\natexlab{}.
\newblock \showarticletitle{Use cases of lossy compression for floating-point
  data in scientific data sets}.
\newblock \bibinfo{journal}{\emph{The International Journal of High Performance
  Computing Applications}} (\bibinfo{year}{2019}).
\newblock


\bibitem[\protect\citeauthoryear{Chen, Tian, Beaver, Freeman, Yan, Wang, and
  Tao}{Chen et~al\mbox{.}}{2024}]%
        {chen2024fcbench}
\bibfield{author}{\bibinfo{person}{Xinyu Chen}, \bibinfo{person}{Jiannan Tian},
  \bibinfo{person}{Ian Beaver}, \bibinfo{person}{Cynthia Freeman},
  \bibinfo{person}{Yan Yan}, \bibinfo{person}{Jianguo Wang}, {and}
  \bibinfo{person}{Dingwen Tao}.} \bibinfo{year}{2024}\natexlab{}.
\newblock \showarticletitle{FCBench: Cross-Domain Benchmarking of Lossless
  Compression for Floating-Point Data}.
\newblock \bibinfo{journal}{\emph{Proceedings of the VLDB Endowment}}
  \bibinfo{volume}{17}, \bibinfo{number}{6} (\bibinfo{year}{2024}),
  \bibinfo{pages}{1418--1431}.
\newblock


\bibitem[\protect\citeauthoryear{Cook, Cabot, and Miller}{Cook
  et~al\mbox{.}}{2004}]%
        {miranda}
\bibfield{author}{\bibinfo{person}{Andrew~W. Cook}, \bibinfo{person}{William
  Cabot}, {and} \bibinfo{person}{Paul~L. Miller}.}
  \bibinfo{year}{2004}\natexlab{}.
\newblock \showarticletitle{The Mixing Transition in {R}ayleigh-{T}aylor
  Instability}.
\newblock \bibinfo{journal}{\emph{Journal of Fluid Mechanics}}
  \bibinfo{volume}{511} (\bibinfo{year}{2004}), \bibinfo{pages}{333--362}.
\newblock
\urldef\tempurl%
\url{https://doi.org/10.1017/S0022112004009681}
\showDOI{\tempurl}


\bibitem[\protect\citeauthoryear{{cuZFP}}{{cuZFP}}{2023}]%
        {cuZFP}
\bibfield{author}{\bibinfo{person}{{cuZFP}}.} \bibinfo{year}{2023}\natexlab{}.
\newblock
  \bibinfo{howpublished}{\url{https://github.com/LLNL/zfp/tree/develop/src/cuda_zfp}}.
\newblock
\newblock
\shownote{Online.}


\bibitem[\protect\citeauthoryear{Davis, Efstathiou, Frenk, and White}{Davis
  et~al\mbox{.}}{1985}]%
        {davis1985evolution}
\bibfield{author}{\bibinfo{person}{Marc Davis}, \bibinfo{person}{George
  Efstathiou}, \bibinfo{person}{Carlos~S Frenk}, {and}
  \bibinfo{person}{Simon~DM White}.} \bibinfo{year}{1985}\natexlab{}.
\newblock \showarticletitle{The evolution of large-scale structure in a
  universe dominated by cold dark matter}.
\newblock \bibinfo{journal}{\emph{The Astrophysical Journal}}
  \bibinfo{volume}{292} (\bibinfo{year}{1985}), \bibinfo{pages}{371--394}.
\newblock


\bibitem[\protect\citeauthoryear{Deutsch}{Deutsch}{1996}]%
        {gzip}
\bibfield{author}{\bibinfo{person}{Peter Deutsch}.}
  \bibinfo{year}{1996}\natexlab{}.
\newblock \bibinfo{booktitle}{\emph{GZIP file format specification version
  4.3}}.
\newblock \bibinfo{type}{{T}echnical {R}eport}.
\newblock


\bibitem[\protect\citeauthoryear{Di and Cappello}{Di and Cappello}{2016}]%
        {di2016fast}
\bibfield{author}{\bibinfo{person}{Sheng Di} {and} \bibinfo{person}{Franck
  Cappello}.} \bibinfo{year}{2016}\natexlab{}.
\newblock \showarticletitle{Fast error-bounded lossy {HPC} data compression
  with {SZ}}. In \bibinfo{booktitle}{\emph{2016 IEEE International Parallel and
  Distributed Processing Symposium}}. IEEE, \bibinfo{pages}{730--739}.
\newblock


\bibitem[\protect\citeauthoryear{Di, Liu, Zhao, Liang, Underwood, Zhang, Shah,
  Huang, Huang, Yu, et~al\mbox{.}}{Di et~al\mbox{.}}{2024}]%
        {di2024survey}
\bibfield{author}{\bibinfo{person}{Sheng Di}, \bibinfo{person}{Jinyang Liu},
  \bibinfo{person}{Kai Zhao}, \bibinfo{person}{Xin Liang},
  \bibinfo{person}{Robert Underwood}, \bibinfo{person}{Zhaorui Zhang},
  \bibinfo{person}{Milan Shah}, \bibinfo{person}{Yafan Huang},
  \bibinfo{person}{Jiajun Huang}, \bibinfo{person}{Xiaodong Yu},
  {et~al\mbox{.}}} \bibinfo{year}{2024}\natexlab{}.
\newblock \showarticletitle{A Survey on Error-Bounded Lossy Compression for
  Scientific Datasets}.
\newblock \bibinfo{journal}{\emph{arXiv preprint arXiv:2404.02840}}
  (\bibinfo{year}{2024}).
\newblock


\bibitem[\protect\citeauthoryear{Fang, Wang, Jin, Koziol, Zhang, Guan, Byna,
  Krishnamoorthy, and Tao}{Fang et~al\mbox{.}}{2021}]%
        {ffis}
\bibfield{author}{\bibinfo{person}{Bo Fang}, \bibinfo{person}{Daoce Wang},
  \bibinfo{person}{Sian Jin}, \bibinfo{person}{Quincey Koziol},
  \bibinfo{person}{Zhao Zhang}, \bibinfo{person}{Qiang Guan},
  \bibinfo{person}{Surendra Byna}, \bibinfo{person}{Sriram Krishnamoorthy},
  {and} \bibinfo{person}{Dingwen Tao}.} \bibinfo{year}{2021}\natexlab{}.
\newblock \showarticletitle{Characterizing Impacts of Storage Faults on HPC
  Applications: A Methodology and Insights}. \bibinfo{pages}{409--420}.
\newblock
\urldef\tempurl%
\url{https://doi.org/10.1109/Cluster48925.2021.00048}
\showDOI{\tempurl}


\bibitem[\protect\citeauthoryear{Fedeli, Huebl, Boillod-Cerneux, Clark, Gott,
  Hillairet, Jaure, Leblanc, Lehe, Myers, Piechurski, Sato, Zaim, Zhang, Vay,
  and Vincenti}{Fedeli et~al\mbox{.}}{2022}]%
        {warpx}
\bibfield{author}{\bibinfo{person}{L. Fedeli}, \bibinfo{person}{A. Huebl},
  \bibinfo{person}{F. Boillod-Cerneux}, \bibinfo{person}{T. Clark},
  \bibinfo{person}{K. Gott}, \bibinfo{person}{C. Hillairet},
  \bibinfo{person}{S. Jaure}, \bibinfo{person}{A. Leblanc}, \bibinfo{person}{R.
  Lehe}, \bibinfo{person}{A. Myers}, \bibinfo{person}{C. Piechurski},
  \bibinfo{person}{M. Sato}, \bibinfo{person}{N. Zaim}, \bibinfo{person}{W.
  Zhang}, \bibinfo{person}{J. Vay}, {and} \bibinfo{person}{H. Vincenti}.}
  \bibinfo{year}{2022}\natexlab{}.
\newblock \showarticletitle{Pushing the Frontier in the Design of Laser-Based
  Electron Accelerators with Groundbreaking Mesh-Refined Particle-In-Cell
  Simulations on Exascale-Class Supercomputers}. In
  \bibinfo{booktitle}{\emph{SC22: International Conference for High Performance
  Computing, Networking, Storage and Analysis}}. \bibinfo{publisher}{IEEE
  Computer Society}, \bibinfo{address}{Los Alamitos, CA, USA},
  \bibinfo{pages}{1--12}.
\newblock
\urldef\tempurl%
\url{https://doi.org/10.1109/SC41404.2022.00008}
\showDOI{\tempurl}


\bibitem[\protect\citeauthoryear{Gong, Chen, Whitney, Liang, Reshniak,
  Banerjee, Lee, Rangarajan, Wan, Vidal, Liu, Gainaru, Podhorszki, Archibald,
  Ranka, and Klasky}{Gong et~al\mbox{.}}{2023}]%
        {gong2023mgard}
\bibfield{author}{\bibinfo{person}{Qian Gong}, \bibinfo{person}{Jieyang Chen},
  \bibinfo{person}{Ben Whitney}, \bibinfo{person}{Xin Liang},
  \bibinfo{person}{Viktor Reshniak}, \bibinfo{person}{Tania Banerjee},
  \bibinfo{person}{Jaemoon Lee}, \bibinfo{person}{Anand Rangarajan},
  \bibinfo{person}{Lipeng Wan}, \bibinfo{person}{Nicolas Vidal},
  \bibinfo{person}{Qing Liu}, \bibinfo{person}{Ana Gainaru},
  \bibinfo{person}{Norbert Podhorszki}, \bibinfo{person}{Richard Archibald},
  \bibinfo{person}{Sanjay Ranka}, {and} \bibinfo{person}{Scott Klasky}.}
  \bibinfo{year}{2023}\natexlab{}.
\newblock \showarticletitle{{MGARD}: A multigrid framework for
  high-performance, error-controlled data compression and refactoring}.
\newblock \bibinfo{journal}{\emph{SoftwareX}}  \bibinfo{volume}{24}
  (\bibinfo{year}{2023}), \bibinfo{pages}{101590}.
\newblock


\bibitem[\protect\citeauthoryear{Guo, Li, Daughton, and Liu}{Guo
  et~al\mbox{.}}{2014}]%
        {magnetic_reconnection}
\bibfield{author}{\bibinfo{person}{Fan Guo}, \bibinfo{person}{Hui Li},
  \bibinfo{person}{William Daughton}, {and} \bibinfo{person}{Yi-Hsin Liu}.}
  \bibinfo{year}{2014}\natexlab{}.
\newblock \showarticletitle{Formation of Hard Power Laws in the Energetic
  Particle Spectra Resulting from Relativistic Magnetic Reconnection}.
\newblock \bibinfo{journal}{\emph{Phys. Rev. Lett.}}  \bibinfo{volume}{113}
  (\bibinfo{date}{oct} \bibinfo{year}{2014}), \bibinfo{pages}{155005}.
\newblock
Issue 15.
\urldef\tempurl%
\url{https://doi.org/10.1103/PhysRevLett.113.155005}
\showDOI{\tempurl}


\bibitem[\protect\citeauthoryear{Huang, Di, Li, and Cappello}{Huang
  et~al\mbox{.}}{2024}]%
        {huang2024cuszp2}
\bibfield{author}{\bibinfo{person}{Yafan Huang}, \bibinfo{person}{Sheng Di},
  \bibinfo{person}{Guanpeng Li}, {and} \bibinfo{person}{Franck Cappello}.}
  \bibinfo{year}{2024}\natexlab{}.
\newblock \showarticletitle{cuSZp2: A GPU Lossy Compressor with Extreme
  Throughput and Optimized Compression Ratio}. In
  \bibinfo{booktitle}{\emph{SC24: International Conference for High Performance
  Computing, Networking, Storage and Analysis}}. IEEE, \bibinfo{pages}{1--18}.
\newblock


\bibitem[\protect\citeauthoryear{Huebl, Widera, Schmitt, Matthes, Podhorszki,
  Choi, Klasky, and Bussmann}{Huebl et~al\mbox{.}}{2017}]%
        {Huebl2017}
\bibfield{author}{\bibinfo{person}{Axel Huebl}, \bibinfo{person}{Ren{\'e}
  Widera}, \bibinfo{person}{Felix Schmitt}, \bibinfo{person}{Alexander
  Matthes}, \bibinfo{person}{Norbert Podhorszki}, \bibinfo{person}{Jong~Youl
  Choi}, \bibinfo{person}{Scott Klasky}, {and} \bibinfo{person}{Michael
  Bussmann}.} \bibinfo{year}{2017}\natexlab{}.
\newblock \showarticletitle{On the Scalability of Data Reduction Techniques in
  Current and Upcoming HPC Systems from an Application Perspective}. In
  \bibinfo{booktitle}{\emph{High Performance Computing}},
  \bibfield{editor}{\bibinfo{person}{Julian~M. Kunkel}, \bibinfo{person}{Rio
  Yokota}, \bibinfo{person}{Michela Taufer}, {and} \bibinfo{person}{John
  Shalf}} (Eds.). \bibinfo{publisher}{Springer International Publishing},
  \bibinfo{address}{Cham}, \bibinfo{pages}{15--29}.
\newblock
\urldef\tempurl%
\url{https://doi.org/10.1007/978-3-319-67630-2_2}
\showDOI{\tempurl}


\bibitem[\protect\citeauthoryear{Jia, Xie, Lu, Wang, Feng, Zhang, Sun, Lin,
  Zhang, Liu, and Tao}{Jia et~al\mbox{.}}{2024b}]%
        {Jia2024SDP4Bit}
\bibfield{author}{\bibinfo{person}{Jinda Jia}, \bibinfo{person}{Cong Xie},
  \bibinfo{person}{Hanlin Lu}, \bibinfo{person}{Daoce Wang},
  \bibinfo{person}{Hao Feng}, \bibinfo{person}{Chengming Zhang},
  \bibinfo{person}{Baixi Sun}, \bibinfo{person}{Haibin Lin},
  \bibinfo{person}{Zhi Zhang}, \bibinfo{person}{Xin Liu}, {and}
  \bibinfo{person}{Dingwen Tao}.} \bibinfo{year}{2024}\natexlab{b}.
\newblock \showarticletitle{SDP4Bit: Toward 4-bit Communication Quantization in
  Sharded Data Parallelism for LLM Training}. In
  \bibinfo{booktitle}{\emph{Advances in Neural Information Processing
  Systems}}, \bibfield{editor}{\bibinfo{person}{A.~Globerson},
  \bibinfo{person}{L.~Mackey}, \bibinfo{person}{D.~Belgrave},
  \bibinfo{person}{A.~Fan}, \bibinfo{person}{U.~Paquet},
  \bibinfo{person}{J.~Tomczak}, {and} \bibinfo{person}{C.~Zhang}} (Eds.),
  Vol.~\bibinfo{volume}{37}. \bibinfo{publisher}{Curran Associates, Inc.},
  \bibinfo{pages}{8734--8759}.
\newblock
\urldef\tempurl%
\url{https://proceedings.neurips.cc/paper_files/paper/2024/file/10826a1a80f816ea98d559d7c7a97973-Paper-Conference.pdf}
\showURL{%
\tempurl}


\bibitem[\protect\citeauthoryear{Jia, Hu, Liu, Zhang, Wang, Liu, Niu,
  Kalafatis, Huang, Jin, et~al\mbox{.}}{Jia et~al\mbox{.}}{2024a}]%
        {jia2024neurlz}
\bibfield{author}{\bibinfo{person}{Wenqi Jia}, \bibinfo{person}{Zhewen Hu},
  \bibinfo{person}{Youyuan Liu}, \bibinfo{person}{Boyuan Zhang},
  \bibinfo{person}{Jinzhen Wang}, \bibinfo{person}{Jinyang Liu},
  \bibinfo{person}{Wei Niu}, \bibinfo{person}{Stavros Kalafatis},
  \bibinfo{person}{Junzhou Huang}, \bibinfo{person}{Sian Jin}, {et~al\mbox{.}}}
  \bibinfo{year}{2024}\natexlab{a}.
\newblock \showarticletitle{NeurLZ: An Online Neural Learning-Based Method to
  Enhance Scientific Lossy Compression}.
\newblock \bibinfo{journal}{\emph{arXiv preprint arXiv:2409.05785}}
  (\bibinfo{year}{2024}).
\newblock


\bibitem[\protect\citeauthoryear{Jin, Di, Vivien, Wang, Robert, Tao, and
  Cappello}{Jin et~al\mbox{.}}{2024}]%
        {jin2024concealing}
\bibfield{author}{\bibinfo{person}{Sian Jin}, \bibinfo{person}{Sheng Di},
  \bibinfo{person}{Fr{\'e}d{\'e}ric Vivien}, \bibinfo{person}{Daoce Wang},
  \bibinfo{person}{Yves Robert}, \bibinfo{person}{Dingwen Tao}, {and}
  \bibinfo{person}{Franck Cappello}.} \bibinfo{year}{2024}\natexlab{}.
\newblock \showarticletitle{Concealing Compression-accelerated I/O for HPC
  Applications through In Situ Task Scheduling}. In
  \bibinfo{booktitle}{\emph{EuroSys 2024}}.
\newblock


\bibitem[\protect\citeauthoryear{Li, Lindstrom, and Clyne}{Li
  et~al\mbox{.}}{2023}]%
        {sperr}
\bibfield{author}{\bibinfo{person}{Shaomeng Li}, \bibinfo{person}{Peter
  Lindstrom}, {and} \bibinfo{person}{John Clyne}.}
  \bibinfo{year}{2023}\natexlab{}.
\newblock \showarticletitle{Lossy Scientific Data Compression With SPERR}. In
  \bibinfo{booktitle}{\emph{2023 IEEE International Parallel and Distributed
  Processing Symposium (IPDPS)}}. \bibinfo{pages}{1007--1017}.
\newblock
\urldef\tempurl%
\url{https://doi.org/10.1109/IPDPS54959.2023.00104}
\showDOI{\tempurl}


\bibitem[\protect\citeauthoryear{Liang, Di, Tao, Li, Li, Guo, Chen, and
  Cappello}{Liang et~al\mbox{.}}{2018}]%
        {sz18}
\bibfield{author}{\bibinfo{person}{Xin Liang}, \bibinfo{person}{Sheng Di},
  \bibinfo{person}{Dingwen Tao}, \bibinfo{person}{Sihuan Li},
  \bibinfo{person}{Shaomeng Li}, \bibinfo{person}{Hanqi Guo},
  \bibinfo{person}{Zizhong Chen}, {and} \bibinfo{person}{Franck Cappello}.}
  \bibinfo{year}{2018}\natexlab{}.
\newblock \showarticletitle{Error-controlled lossy compression optimized for
  high compression ratios of scientific datasets}. In
  \bibinfo{booktitle}{\emph{2018 IEEE International Conference on Big Data}}.
  IEEE, \bibinfo{pages}{438--447}.
\newblock


\bibitem[\protect\citeauthoryear{Liang, Gong, Chen, Whitney, Wan, Liu, Pugmire,
  Archibald, Podhorszki, and Klasky}{Liang et~al\mbox{.}}{2021}]%
        {liang2021error}
\bibfield{author}{\bibinfo{person}{Xin Liang}, \bibinfo{person}{Qian Gong},
  \bibinfo{person}{Jieyang Chen}, \bibinfo{person}{Ben Whitney},
  \bibinfo{person}{Lipeng Wan}, \bibinfo{person}{Qing Liu},
  \bibinfo{person}{David Pugmire}, \bibinfo{person}{Rick Archibald},
  \bibinfo{person}{Norbert Podhorszki}, {and} \bibinfo{person}{Scott Klasky}.}
  \bibinfo{year}{2021}\natexlab{}.
\newblock \showarticletitle{Error-controlled, progressive, and adaptable
  retrieval of scientific data with multilevel decomposition}. In
  \bibinfo{booktitle}{\emph{Proceedings of the International Conference for
  High Performance Computing, Networking, Storage and Analysis}}.
  \bibinfo{pages}{1--13}.
\newblock


\bibitem[\protect\citeauthoryear{Lindstrom}{Lindstrom}{2014}]%
        {zfp}
\bibfield{author}{\bibinfo{person}{Peter Lindstrom}.}
  \bibinfo{year}{2014}\natexlab{}.
\newblock \showarticletitle{Fixed-rate compressed floating-point arrays}.
\newblock \bibinfo{journal}{\emph{IEEE Transactions on Visualization and
  Computer Graphics}} \bibinfo{volume}{20}, \bibinfo{number}{12}
  (\bibinfo{year}{2014}), \bibinfo{pages}{2674--2683}.
\newblock


\bibitem[\protect\citeauthoryear{Liu, Tian, Wu, Di, Zhang, Underwood, Huang,
  Huang, Zhao, Li, et~al\mbox{.}}{Liu et~al\mbox{.}}{2024}]%
        {liu2024cusz}
\bibfield{author}{\bibinfo{person}{Jinyang Liu}, \bibinfo{person}{Jiannan
  Tian}, \bibinfo{person}{Shixun Wu}, \bibinfo{person}{Sheng Di},
  \bibinfo{person}{Boyuan Zhang}, \bibinfo{person}{Robert Underwood},
  \bibinfo{person}{Yafan Huang}, \bibinfo{person}{Jiajun Huang},
  \bibinfo{person}{Kai Zhao}, \bibinfo{person}{Guanpeng Li}, {et~al\mbox{.}}}
  \bibinfo{year}{2024}\natexlab{}.
\newblock \showarticletitle{CUSZ-i: High-Ratio Scientific Lossy Compression on
  GPUs with Optimized Multi-Level Interpolation}. In
  \bibinfo{booktitle}{\emph{SC24: International Conference for High Performance
  Computing, Networking, Storage and Analysis}}. IEEE, \bibinfo{pages}{1--15}.
\newblock


\bibitem[\protect\citeauthoryear{Lu, Liu, He, Luo, Suchyta, Choi, Podhorszki,
  Klasky, Wolf, Liu, et~al\mbox{.}}{Lu et~al\mbox{.}}{2018}]%
        {lu2018understanding}
\bibfield{author}{\bibinfo{person}{Tao Lu}, \bibinfo{person}{Qing Liu},
  \bibinfo{person}{Xubin He}, \bibinfo{person}{Huizhang Luo},
  \bibinfo{person}{Eric Suchyta}, \bibinfo{person}{Jong Choi},
  \bibinfo{person}{Norbert Podhorszki}, \bibinfo{person}{Scott Klasky},
  \bibinfo{person}{Mathew Wolf}, \bibinfo{person}{Tong Liu}, {et~al\mbox{.}}}
  \bibinfo{year}{2018}\natexlab{}.
\newblock \showarticletitle{Understanding and modeling lossy compression
  schemes on {HPC} scientific data}. In \bibinfo{booktitle}{\emph{2018 IEEE
  International Parallel and Distributed Processing Symposium}}. IEEE,
  \bibinfo{pages}{348--357}.
\newblock


\bibitem[\protect\citeauthoryear{Luo, Huang, Liu, Qiao, Jiang, Bi, Yuan, Zhou,
  Wang, and Qin}{Luo et~al\mbox{.}}{2019}]%
        {luo2019identifying}
\bibfield{author}{\bibinfo{person}{Huizhang Luo}, \bibinfo{person}{Dan Huang},
  \bibinfo{person}{Qing Liu}, \bibinfo{person}{Zhenbo Qiao},
  \bibinfo{person}{Hong Jiang}, \bibinfo{person}{Jing Bi},
  \bibinfo{person}{Haitao Yuan}, \bibinfo{person}{Mengchu Zhou},
  \bibinfo{person}{Jinzhen Wang}, {and} \bibinfo{person}{Zhenlu Qin}.}
  \bibinfo{year}{2019}\natexlab{}.
\newblock \showarticletitle{Identifying Latent Reduced Models to Precondition
  Lossy Compression}. In \bibinfo{booktitle}{\emph{2019 IEEE International
  Parallel and Distributed Processing Symposium}}. IEEE.
\newblock


\bibitem[\protect\citeauthoryear{Magri and Lindstrom}{Magri and
  Lindstrom}{2023}]%
        {magri2023general}
\bibfield{author}{\bibinfo{person}{Victor~AP Magri} {and}
  \bibinfo{person}{Peter Lindstrom}.} \bibinfo{year}{2023}\natexlab{}.
\newblock \showarticletitle{A general framework for progressive data
  compression and retrieval}.
\newblock \bibinfo{journal}{\emph{IEEE Transactions on Visualization and
  Computer Graphics}} \bibinfo{volume}{30}, \bibinfo{number}{1}
  (\bibinfo{year}{2023}), \bibinfo{pages}{1358--1368}.
\newblock


\bibitem[\protect\citeauthoryear{{NYX simulation}}{{NYX simulation}}{2019}]%
        {nyx}
\bibfield{author}{\bibinfo{person}{{NYX simulation}}.}
  \bibinfo{year}{2019}\natexlab{}.
\newblock \bibinfo{howpublished}{\url{https://amrex-astro.github.io/Nyx/}}.
\newblock
\newblock
\shownote{Online.}


\bibitem[\protect\citeauthoryear{{Oak Ridge Leadership Computing
  Facility}}{{Oak Ridge Leadership Computing Facility}}{2023}]%
        {warpx-gordon}
\bibfield{author}{\bibinfo{person}{{Oak Ridge Leadership Computing Facility}}.}
  \bibinfo{year}{2023}\natexlab{}.
\newblock \bibinfo{booktitle}{\emph{{WarpX, granted early access to the
  exascale supercomputer Frontier, receives the high-performance computing
  world’s highest honor}}}.
\newblock
\urldef\tempurl%
\url{https://www.olcf.ornl.gov/2022/11/17/plasma-simulation-code-wins-2022-acm-gordon-bell-prize/}
\showURL{%
\tempurl}
\newblock
\shownote{Online.}


\bibitem[\protect\citeauthoryear{Sun, Liu, Pauloski, Tian, Jia, Wang, Zhang,
  Zheng, Di, Jin, Zhang, Yu, Iskra, Beckman, Tan, and Tao}{Sun
  et~al\mbox{.}}{2025}]%
        {sun2025compso}
\bibfield{author}{\bibinfo{person}{Baixi Sun}, \bibinfo{person}{Weijin Liu},
  \bibinfo{person}{J.~Gregory Pauloski}, \bibinfo{person}{Jiannan Tian},
  \bibinfo{person}{Jinda Jia}, \bibinfo{person}{Daoce Wang},
  \bibinfo{person}{Boyuan Zhang}, \bibinfo{person}{Mingkai Zheng},
  \bibinfo{person}{Sheng Di}, \bibinfo{person}{Sian Jin}, \bibinfo{person}{Zhao
  Zhang}, \bibinfo{person}{Xiaodong Yu}, \bibinfo{person}{Kamil~A. Iskra},
  \bibinfo{person}{Pete Beckman}, \bibinfo{person}{Guangming Tan}, {and}
  \bibinfo{person}{Dingwen Tao}.} \bibinfo{year}{2025}\natexlab{}.
\newblock \showarticletitle{COMPSO: Optimizing Gradient Compression for
  Distributed Training with Second-Order Optimizers}. In
  \bibinfo{booktitle}{\emph{Proceedings of the 30th ACM SIGPLAN Annual
  Symposium on Principles and Practice of Parallel Programming}} (Las Vegas,
  NV, USA) \emph{(\bibinfo{series}{PPoPP '25})}.
  \bibinfo{publisher}{Association for Computing Machinery},
  \bibinfo{address}{New York, NY, USA}, \bibinfo{pages}{212–224}.
\newblock
\showISBNx{9798400714436}
\urldef\tempurl%
\url{https://doi.org/10.1145/3710848.3710852}
\showDOI{\tempurl}


\bibitem[\protect\citeauthoryear{Tao, Di, Chen, and Cappello}{Tao
  et~al\mbox{.}}{2017}]%
        {sz17}
\bibfield{author}{\bibinfo{person}{Dingwen Tao}, \bibinfo{person}{Sheng Di},
  \bibinfo{person}{Zizhong Chen}, {and} \bibinfo{person}{Franck Cappello}.}
  \bibinfo{year}{2017}\natexlab{}.
\newblock \showarticletitle{Significantly improving lossy compression for
  scientific data sets based on multidimensional prediction and
  error-controlled quantization}. In \bibinfo{booktitle}{\emph{2017 IEEE
  International Parallel and Distributed Processing Symposium}}. IEEE,
  \bibinfo{pages}{1129--1139}.
\newblock


\bibitem[\protect\citeauthoryear{Tian, Di, Yu, Rivera, Zhao, Jin, Feng, Liang,
  Tao, and Cappello}{Tian et~al\mbox{.}}{2021}]%
        {tian2021optimizing}
\bibfield{author}{\bibinfo{person}{Jiannan Tian}, \bibinfo{person}{Sheng Di},
  \bibinfo{person}{Xiaodong Yu}, \bibinfo{person}{Cody Rivera},
  \bibinfo{person}{Kai Zhao}, \bibinfo{person}{Sian Jin},
  \bibinfo{person}{Yunhe Feng}, \bibinfo{person}{Xin Liang},
  \bibinfo{person}{Dingwen Tao}, {and} \bibinfo{person}{Franck Cappello}.}
  \bibinfo{year}{2021}\natexlab{}.
\newblock \showarticletitle{Optimizing error-bounded lossy compression for
  scientific data on GPUs}. In \bibinfo{booktitle}{\emph{2021 IEEE
  International Conference on Cluster Computing (CLUSTER)}}. IEEE,
  \bibinfo{pages}{283--293}.
\newblock


\bibitem[\protect\citeauthoryear{Tian, Di, Zhao, Rivera, Fulp, Underwood, Jin,
  Liang, Calhoun, Tao, and Cappello}{Tian et~al\mbox{.}}{2020}]%
        {tian2020cusz}
\bibfield{author}{\bibinfo{person}{Jiannan Tian}, \bibinfo{person}{Sheng Di},
  \bibinfo{person}{Kai Zhao}, \bibinfo{person}{Cody Rivera},
  \bibinfo{person}{Megan~Hickman Fulp}, \bibinfo{person}{Robert Underwood},
  \bibinfo{person}{Sian Jin}, \bibinfo{person}{Xin Liang}, \bibinfo{person}{Jon
  Calhoun}, \bibinfo{person}{Dingwen Tao}, {and} \bibinfo{person}{Franck
  Cappello}.} \bibinfo{year}{2020}\natexlab{}.
\newblock \showarticletitle{cuSZ: An Efficient GPU-Based Error-Bounded Lossy
  Compression Framework for Scientific Data}.
\newblock  (\bibinfo{year}{2020}), \bibinfo{pages}{3--15}.
\newblock


\bibitem[\protect\citeauthoryear{Ulmer, Angelini, Fekete, Kohlhammer, and
  May}{Ulmer et~al\mbox{.}}{2024}]%
        {Progressive2}
\bibfield{author}{\bibinfo{person}{Alex Ulmer}, \bibinfo{person}{Marco
  Angelini}, \bibinfo{person}{Jean-Daniel Fekete}, \bibinfo{person}{Jörn
  Kohlhammer}, {and} \bibinfo{person}{Thorsten May}.}
  \bibinfo{year}{2024}\natexlab{}.
\newblock \showarticletitle{A Survey on Progressive Visualization}.
\newblock \bibinfo{journal}{\emph{IEEE Transactions on Visualization and
  Computer Graphics}} \bibinfo{volume}{30}, \bibinfo{number}{9}
  (\bibinfo{year}{2024}), \bibinfo{pages}{6447--6467}.
\newblock
\urldef\tempurl%
\url{https://doi.org/10.1109/TVCG.2023.3346641}
\showDOI{\tempurl}


\bibitem[\protect\citeauthoryear{Wang}{Wang}{2025}]%
        {wang2025designing}
\bibfield{author}{\bibinfo{person}{Daoce Wang}.}
  \bibinfo{year}{2025}\natexlab{}.
\newblock \emph{\bibinfo{title}{Designing Efficient Data Reduction Approaches
  for Multi-Resolution Simulations on HPC Systems}}.
\newblock \bibinfo{thesistype}{Ph.D. Dissertation}. \bibinfo{school}{Indiana
  University}.
\newblock


\bibitem[\protect\citeauthoryear{Wang, Grosset, Pulido, Athawale, Tian, Zhao,
  Lukić, Huebl, Wang, Ahrens, and Tao}{Wang et~al\mbox{.}}{2024a}]%
        {wang2024mrz}
\bibfield{author}{\bibinfo{person}{Daoce Wang}, \bibinfo{person}{Pascal
  Grosset}, \bibinfo{person}{Jesus Pulido}, \bibinfo{person}{Tushar~M.
  Athawale}, \bibinfo{person}{Jiannan Tian}, \bibinfo{person}{Kai Zhao},
  \bibinfo{person}{Zarija Lukić}, \bibinfo{person}{Axel Huebl},
  \bibinfo{person}{Zhe Wang}, \bibinfo{person}{James Ahrens}, {and}
  \bibinfo{person}{Dingwen Tao}.} \bibinfo{year}{2024}\natexlab{a}.
\newblock \showarticletitle{A High-Quality Workflow for Multi-Resolution
  Scientific Data Reduction and Visualization}. In
  \bibinfo{booktitle}{\emph{SC24: International Conference for High Performance
  Computing, Networking, Storage and Analysis}}. \bibinfo{pages}{1--18}.
\newblock
\urldef\tempurl%
\url{https://doi.org/10.1109/SC41406.2024.00091}
\showDOI{\tempurl}


\bibitem[\protect\citeauthoryear{Wang, Pulido, Grosset, Jin, Tian, Ahrens, and
  Tao}{Wang et~al\mbox{.}}{2022}]%
        {wang2022tac}
\bibfield{author}{\bibinfo{person}{Daoce Wang}, \bibinfo{person}{Jesus Pulido},
  \bibinfo{person}{Pascal Grosset}, \bibinfo{person}{Sian Jin},
  \bibinfo{person}{Jiannan Tian}, \bibinfo{person}{James Ahrens}, {and}
  \bibinfo{person}{Dingwen Tao}.} \bibinfo{year}{2022}\natexlab{}.
\newblock \showarticletitle{TAC: Optimizing Error-Bounded Lossy Compression for
  Three-Dimensional Adaptive Mesh Refinement Simulations}. In
  \bibinfo{booktitle}{\emph{Proceedings of the 31st International Symposium on
  High-Performance Parallel and Distributed Computing}}.
  \bibinfo{pages}{135--147}.
\newblock


\bibitem[\protect\citeauthoryear{Wang, Pulido, Grosset, Jin, Tian, Zhao,
  Ahrens, and Tao}{Wang et~al\mbox{.}}{2024b}]%
        {wang2024tac+}
\bibfield{author}{\bibinfo{person}{Daoce Wang}, \bibinfo{person}{Jesus Pulido},
  \bibinfo{person}{Pascal Grosset}, \bibinfo{person}{Sian Jin},
  \bibinfo{person}{Jiannan Tian}, \bibinfo{person}{Kai Zhao},
  \bibinfo{person}{James Ahrens}, {and} \bibinfo{person}{Dingwen Tao}.}
  \bibinfo{year}{2024}\natexlab{b}.
\newblock \showarticletitle{TAC+: Optimizing Error-Bounded Lossy Compression
  for 3D AMR Simulations}.
\newblock \bibinfo{journal}{\emph{IEEE Transactions on Parallel and Distributed
  Systems}} \bibinfo{volume}{35}, \bibinfo{number}{3} (\bibinfo{year}{2024}),
  \bibinfo{pages}{421--438}.
\newblock


\bibitem[\protect\citeauthoryear{Wang, Pulido, Grosset, Tian, Ahrens, and
  Tao}{Wang et~al\mbox{.}}{2023a}]%
        {wang2023amrvis}
\bibfield{author}{\bibinfo{person}{Daoce Wang}, \bibinfo{person}{Jesus Pulido},
  \bibinfo{person}{Pascal Grosset}, \bibinfo{person}{Jiannan Tian},
  \bibinfo{person}{James Ahrens}, {and} \bibinfo{person}{Dingwen Tao}.}
  \bibinfo{year}{2023}\natexlab{a}.
\newblock \showarticletitle{Analyzing impact of data reduction techniques on
  visualization for amr applications using amrex framework}. In
  \bibinfo{booktitle}{\emph{Proceedings of the SC'23 Workshops of The
  International Conference on High Performance Computing, Network, Storage, and
  Analysis}}. \bibinfo{pages}{263--271}.
\newblock


\bibitem[\protect\citeauthoryear{Wang, Pulido, Grosset, Tian, Jin, Tang,
  Sexton, Di, Zhao, Fang, et~al\mbox{.}}{Wang et~al\mbox{.}}{2023b}]%
        {amric}
\bibfield{author}{\bibinfo{person}{Daoce Wang}, \bibinfo{person}{Jesus Pulido},
  \bibinfo{person}{Pascal Grosset}, \bibinfo{person}{Jiannan Tian},
  \bibinfo{person}{Sian Jin}, \bibinfo{person}{Houjun Tang},
  \bibinfo{person}{Jean Sexton}, \bibinfo{person}{Sheng Di},
  \bibinfo{person}{Kai Zhao}, \bibinfo{person}{Bo Fang}, {et~al\mbox{.}}}
  \bibinfo{year}{2023}\natexlab{b}.
\newblock \showarticletitle{AMRIC: A Novel In Situ Lossy Compression Framework
  for Efficient I/O in Adaptive Mesh Refinement Applications}. In
  \bibinfo{booktitle}{\emph{Proceedings of the International Conference for
  High Performance Computing, Networking, Storage and Analysis}}.
  \bibinfo{pages}{1--15}.
\newblock


\bibitem[\protect\citeauthoryear{Wang, Bovik, Sheikh, and Simoncelli}{Wang
  et~al\mbox{.}}{2004}]%
        {ssim}
\bibfield{author}{\bibinfo{person}{Zhou Wang}, \bibinfo{person}{Alan~C Bovik},
  \bibinfo{person}{Hamid~R Sheikh}, {and} \bibinfo{person}{Eero~P Simoncelli}.}
  \bibinfo{year}{2004}\natexlab{}.
\newblock \showarticletitle{Image quality assessment: from error visibility to
  structural similarity}.
\newblock \bibinfo{journal}{\emph{IEEE transactions on image processing}}
  \bibinfo{volume}{13}, \bibinfo{number}{4} (\bibinfo{year}{2004}),
  \bibinfo{pages}{600--612}.
\newblock


\bibitem[\protect\citeauthoryear{Wu, Gong, Chen, Liu, Podhorszki, Liang, and
  Klasky}{Wu et~al\mbox{.}}{2024}]%
        {wu2024error}
\bibfield{author}{\bibinfo{person}{Xuan Wu}, \bibinfo{person}{Qian Gong},
  \bibinfo{person}{Jieyang Chen}, \bibinfo{person}{Qing Liu},
  \bibinfo{person}{Norbert Podhorszki}, \bibinfo{person}{Xin Liang}, {and}
  \bibinfo{person}{Scott Klasky}.} \bibinfo{year}{2024}\natexlab{}.
\newblock \showarticletitle{Error-controlled Progressive Retrieval of
  Scientific Data under Derivable Quantities of Interest}. In
  \bibinfo{booktitle}{\emph{SC24: International Conference for High Performance
  Computing, Networking, Storage and Analysis}}. IEEE, \bibinfo{pages}{1--16}.
\newblock


\bibitem[\protect\citeauthoryear{Yang, Di, Zhang, Li, Li, Huang, Liu, Cappello,
  and Zhao}{Yang et~al\mbox{.}}{2025}]%
        {yang2025ipcomp}
\bibfield{author}{\bibinfo{person}{Zhuoxun Yang}, \bibinfo{person}{Sheng Di},
  \bibinfo{person}{Longtao Zhang}, \bibinfo{person}{Ruoyu Li},
  \bibinfo{person}{Ximiao Li}, \bibinfo{person}{Jiajun Huang},
  \bibinfo{person}{Jinyang Liu}, \bibinfo{person}{Franck Cappello}, {and}
  \bibinfo{person}{Kai Zhao}.} \bibinfo{year}{2025}\natexlab{}.
\newblock \bibinfo{title}{IPComp: Interpolation Based Progressive Lossy
  Compression for Scientific Applications}.
\newblock
\newblock
\showeprint[arxiv]{2502.04093}~[cs.DC]
\urldef\tempurl%
\url{https://arxiv.org/abs/2502.04093}
\showURL{%
\tempurl}


\bibitem[\protect\citeauthoryear{Zgraggen, Galakatos, Crotty, Fekete, and
  Kraska}{Zgraggen et~al\mbox{.}}{2017}]%
        {Progressive1}
\bibfield{author}{\bibinfo{person}{Emanuel Zgraggen}, \bibinfo{person}{Alex
  Galakatos}, \bibinfo{person}{Andrew Crotty}, \bibinfo{person}{Jean-Daniel
  Fekete}, {and} \bibinfo{person}{Tim Kraska}.}
  \bibinfo{year}{2017}\natexlab{}.
\newblock \showarticletitle{How Progressive Visualizations Affect Exploratory
  Analysis}.
\newblock \bibinfo{journal}{\emph{IEEE Transactions on Visualization and
  Computer Graphics}} \bibinfo{volume}{23}, \bibinfo{number}{8}
  (\bibinfo{year}{2017}), \bibinfo{pages}{1977--1987}.
\newblock
\urldef\tempurl%
\url{https://doi.org/10.1109/TVCG.2016.2607714}
\showDOI{\tempurl}


\bibitem[\protect\citeauthoryear{Zhao, Di, Dmitriev, Tonellot, Chen, and
  Cappello}{Zhao et~al\mbox{.}}{2021}]%
        {zhao2021optimizing}
\bibfield{author}{\bibinfo{person}{Kai Zhao}, \bibinfo{person}{Sheng Di},
  \bibinfo{person}{Maxim Dmitriev}, \bibinfo{person}{Thierry-Laurent~D
  Tonellot}, \bibinfo{person}{Zizhong Chen}, {and} \bibinfo{person}{Franck
  Cappello}.} \bibinfo{year}{2021}\natexlab{}.
\newblock \showarticletitle{Optimizing error-bounded lossy compression for
  scientific data by dynamic spline interpolation}. In
  \bibinfo{booktitle}{\emph{2021 IEEE 37th International Conference on Data
  Engineering (ICDE)}}. IEEE, \bibinfo{pages}{1643--1654}.
\newblock


\bibitem[\protect\citeauthoryear{Zheng, Atkinson, Wang, Lee, Patchett, Manno,
  and Grider}{Zheng et~al\mbox{.}}{2024}]%
        {zheng2024vtk}
\bibfield{author}{\bibinfo{person}{Qing Zheng}, \bibinfo{person}{Brian
  Atkinson}, \bibinfo{person}{Daoce Wang}, \bibinfo{person}{Jason Lee},
  \bibinfo{person}{John Patchett}, \bibinfo{person}{Dominic Manno}, {and}
  \bibinfo{person}{Gary Grider}.} \bibinfo{year}{2024}\natexlab{}.
\newblock \showarticletitle{Accelerating Viz Pipelines Using Near-Data
  Computing: An Early Experience}. In \bibinfo{booktitle}{\emph{SC24-W:
  Workshops of the International Conference for High Performance Computing,
  Networking, Storage and Analysis}}. IEEE, \bibinfo{pages}{326--335}.
\newblock


\bibitem[\protect\citeauthoryear{Zstandard}{Zstandard}{2020}]%
        {zstd}
\bibfield{author}{\bibinfo{person}{Zstandard}.}
  \bibinfo{year}{2020}\natexlab{}.
\newblock \bibinfo{howpublished}{\url{http://facebook.github.io/zstd/}}.
\newblock


\end{thebibliography}

\end{document}